\documentclass[preprint,12pt]{elsarticle}

\usepackage{amssymb}
\usepackage{amsmath} 
\usepackage{listings}
\usepackage{xcolor}  
\usepackage{siunitx} 

\DeclareSIUnit\statcoulomb{statC}
\DeclareSIUnit\statvolt{statV}
\DeclareSIUnit\erg{erg}

\colorlet{punct}{red!60!black}
\definecolor{delim}{RGB}{20,105,176}
\colorlet{digit}{magenta!60!black}

\lstdefinelanguage{json}{
    basicstyle=\normalfont\ttfamily,
    numbers=left,
    numberstyle=\scriptsize,
    showstringspaces=false,
    frame=l,
    literate=
     *{0}{{{\color{digit}0}}}{1}
      {1}{{{\color{digit}1}}}{1}
      {2}{{{\color{digit}2}}}{1}
      {3}{{{\color{digit}3}}}{1}
      {4}{{{\color{digit}4}}}{1}
      {5}{{{\color{digit}5}}}{1}
      {6}{{{\color{digit}6}}}{1}
      {7}{{{\color{digit}7}}}{1}
      {8}{{{\color{digit}8}}}{1}
      {9}{{{\color{digit}9}}}{1}
      {:}{{{\color{punct}{:}}}}{1}
      {,}{{{\color{punct}{,}}}}{1}
      {\{}{{{\color{delim}{\{}}}}{1}
      {\}}{{{\color{delim}{\}}}}}{1}
      {[}{{{\color{delim}{[}}}}{1}
      {]}{{{\color{delim}{]}}}}{1},
}

\newcounter{bla}

\journal{Computer Physics Communications}

\begin{document}

\begin{frontmatter}


\title{Afterlive: A performant code for Vlasov-Hybrid simulations}

\author[a]{Patrick Kilian\corref{author}}
\author[a,b,c]{Cedric Schreiner}
\author[a]{Felix Spanier}

\cortext[author] {Corresponding author.\\\textit{E-mail address:} mail@petschge.de}
\address[a]{North-West University, Potchefstroom, South Africa}
\address[b]{Julius-Maximilians-Universit\"at, W\"urzburg, Germany}
\address[c]{Max-Planck-Institute for Solar System Research, G\"ottingen, Germany}

\begin{abstract}
A parallelized implementation of the Vlasov-Hybrid method \cite{Nunn_1993} is
presented. This method is a hybrid between a gridded Eulerian description and
Lagrangian meta-particles. Unlike the Particle-in-Cell method
\cite{Dawson_1983} which simply adds up the contribution of meta-particles,
this method does a reconstruction of the distribution function $f$ in every
time step for each species. This interpolation method combines meta-particles
with different weights in such a way that particles with large weight do not
drown out particles that represent small contributions to the phase space
density.  These core properties allow the use of a much larger range of macro
factors and can thus represent a much larger dynamic range in phase space
density.

The reconstructed phase space density $f$ is used to calculate momenta of the
distribution function such as the charge density $\rho$.
The charge density $\rho$ is also used as input into a spectral solver that
calculates the self-consistent electrostatic field which is used to update the
particles for the next time-step.

Afterlive (\textbf{A} \textbf{F}ourier-based \textbf{T}ool in the
\textbf{E}lectrostatic limit for the \textbf{R}apid \textbf{L}ow-noise
\textbf{I}ntegration of the \textbf{V}lasov \textbf{E}quation) is fully
parallelized using MPI and writes output using parallel HDF5. The input to the
simulation is read from a JSON description that sets the initial particle
distributions as well as domain size and discretization constraints. The
implementation presented here is intentionally limited to one spatial dimension and resolves one or three dimensions in velocity space.
Additional spatial dimensions can be added in a straight forward
way, but make runs computationally even more costly.
\end{abstract}




\end{frontmatter}

{\bf PROGRAM SUMMARY}

\begin{small}
\noindent
{\em Manuscript Title:} Afterlive: A performant code for Vlasov-Hybrid simulations \\
{\em Authors:} Patrick Kilian, Cedric Schreier and Felix Spanier \\
{\em Program Title:} Afterlive (\textbf{A} \textbf{F}ourier-based \textbf{T}ool in the \textbf{E}lectrostatic limit for the \textbf{R}apid \textbf{L}ow-noise \textbf{I}ntegration of the \textbf{V}lasov \textbf{E}quation)                               \\
{\em Journal Reference:}                                      \\ 
{\em Catalogue identifier:}                                   \\ 
{\em Licensing provisions:} GNU Public Licence                \\ 
{\em Programming language:} C++                               \\
{\em Computer:} any workstation or cluster that has a modern C++ compiler (e.g. g++ 4.9 or later) an MPI implementation and the required external libraries \\
{\em Operating system:} Linux / Unix                          \\
{\em RAM:} depending on problem size and number of resolved velocity dimensions between 100 MB and 1 TB; 24 bytes per phase point marker, 24 bytes per grid point and species in phase space, some scratch space \\
{\em Number of processors used:} depending on problem size between one and a few hundred processors 
{\em Classification:} 19.3 Collisionless Plasmas              \\ 
{\em External routines/libraries:} C++ compiler (tested with g++ 4.8, 4.9, 5.3 and 6.3), MPI 1.1 (tested with OpenMPI 1.6.5, 1.8.1, and 1.10.2 and MPICH 3.1), HDF5 with support for parallel I/O(tested with version 1.8.13, 1.8.15, 1.8.16 and 1.10.0), FFTW3 (tested with version 3.3.3, 3.3.4 and 3.3.5), Blitz++ (tested with version 0.10), Jansson (tested with version 2.7.1, 2.7.3, 2.7.5 and 2.9.1) and pkg-config\\
{\em Nature of problem:} Kinetic simulations of collisionless plasma are often done with particle-in-cell codes which also mix Eulerian (fields on a grid) and Lagrangian (freely flowing meta-particles to track particle densities) description. In these codes however computational meta-particles are added up in the deposition scheme. Thus particles are usually equal weights and the noise level is considerable. Purely Eulerian codes must go to great length to yield a stable scheme without excessive numerical diffusion.\\
{\em Solution method:} In this method a Eulerian description of the electrostatic field is also combined with a Lagrangian description of particles, but the distribution function $f$ is reconstructed over the full phase space in a way that avoids drowning out particles with a low phase space weight. This allows for the use of a wide range of macro factors for different tracer particles and an excellent dynamic range in densities. \\
{\em Restrictions:} The current implementation handles one dimension in space and one or three dimensions in velocity only and is applicable to electrostatic scenarios. Using the two extra components of the velocity is computationally costly. \\
{\em Unusual features:} The configuration is stored in JSON files. \\
{\em Running time:} minutes to hours; $10^6$ to $10^7$ marker updates per second and CPU.\\

\end{small}

\section{Introduction}
\label{sec:intro}

This paper provides and discusses a simulation code for the study of
collisionless, electrostatic, one-dimensional plasma.
In this context collisionless means that Coulomb interactions at very short
distances between two particles are rare and don't contribute significantly to
the dynamics of the charged particles compared to the effect of the
electromagnetic fields that are collectively generated by all particles. This
is often the case in astrophysical plasmas as they are usually sufficiently
dilute and warm to have a very large number of particles within a Debye sphere.

A lot of these cases also can very effectively be described as electrostatic.
In this limit only the effect of charge density fluctuations that produce
electric fields are considered, whereas the magnetic fields that are produced
by fluctuations in the current density are ignored. For a plasma with flow
speeds much slower than the speed of light and without large scale structures
in the flow field that produce strong currents this is well justified. The big
advantage of this approach is the possibility to calculate the electric field
directly from the charge density distribution $\rho$. This removes the need to
calculate the coupled time evolution of the electric and magnetic field, which
introduces a severe limitation on the time-step through a CFL condition
involving the spatial discretization scale and the speed of light. The downside
is of course that interactions of the plasma with electromagnetic waves (such as
radio waves) cannot be studied.

Neglecting collisions and magnetic fields generated by the plasma are often
excellent approximations to reality. Restricting the plasma to a single spatial
dimension is much harder to justify. Nevertheless there is a long history and
large number of simulation codes that do exactly that. The reason is that
computer codes are much easier to write, understand and run in this limit.
There are no tedious repetitions of code that reiterate the same idea for
different dimensions or loops over the spatial dimensions mixed into the
essential structure of the code. The limited number of independent dimensions
(essentially $x$, $v_\mathrm{x}$ and $t$) makes the selection and creation of
plots easy. Furthermore the computational effort is usually small and can be
handled on individual machines.
%
%
We therefore present Afterlife in a 1d1v configuration to provide as much
clarity in the code as possible. The user may first get acquainted with the
VHS method and its implementation before performing the transition to the
1d3v configuration that is also supported by the code. This possible configuration is also
the reason for the use of
a MPI-based domain decomposition instead of a perhaps easier to understand
parallelization based on OpenMP. The CPU power and size of main memory necessary for simulations with
multiple dimensions cannot easily be provided by a single machine.
Hybrid parallelization using both OpenMP and MPI has been considered, but is
more complicated and is not necessary due to the good scalability of the code.

Even with highly scalable codes and the largest available supercomputers it is
not feasible to track the motion of each individual particle in the plasma. And
to a large extent we are not even interested in the trajectory of a single
particle but rather in the overall evolution of the system. In the approximation
described above and considering only the 1d1v case for ease of notation the
evolution of the system is given by the Vlasov-Poisson system.

The Vlasov equation gives the time evolution of the phase space density
$f_{\alpha}\left(x, v_\mathrm{x}, t\right)$ of each species $\alpha$ in the
plasma:
\begin{equation}
\frac{\mathrm{d}f_\alpha}{\mathrm{d} t} = \frac{\partial f_\alpha}{\partial t} + v_\mathrm{x} \cdot \frac{\partial f_\alpha}{\partial x} + q_\alpha \, E_\mathrm{x}\left(x, t\right) \cdot \frac{\partial f_\alpha}{\partial v_\mathrm{x}} = 0\quad .
\label{eqn:vlasov}
\end{equation}
The phase space density is a continuous quantity that describes the density of particles as a function of position, velocity and time, ignoring the discrete nature of the particles. Due to the immense number of particles under consideration this is, however, an excellent approximation.

The evolution equation for the space space density is coupled with the Poisson
equation for the electrostatic field
$E_\mathrm{x}\left(x, t\right)$ that depends on the net charge density
$\rho\left(x, t\right)$ that is produced by all species in the plasma:
\begin{equation}
\frac{\partial}{\partial x} E_x\left(x, t\right) = 4 \pi \rho\left(x, t\right) = 4 \pi \, \sum_{\alpha} q_\alpha \int_{-\infty}^{\infty}f\left(x, v_\mathrm{x}, t\right)\, \mathrm{d}v \quad .
\label{eqn:poisson}
\end{equation}

Analytic solutions to this coupled set of equations can only be obtained for
special cases. Nevertheless an analytic approach can be very useful to find
equilibria or properties of small amplitude waves. In many other cases we have
to resort to numerical methods.

Over the decades many different ideas on how to discretize the equations for phase space density and field evolution have
been studied and even more different implementations have be written. This
paper does not aim to cover all the different approaches, but it is worthwhile to
mention a few main lines of thought to give context to the Vlasov-Hybrid
simulation as one possible method among many others, with is own advantages and
drawbacks.

The large and diverse family of Eulerian codes discretizes phase space.
Following Liouville's theorem, the evolution of the phase space density of each species resembles the dynamics of an incompressible fluid. Consequently, all methods from computational fluid dynamics can be used to solve the time evolution by describing the flow of the incompressible fluid on the grid in phase space.
From the phase space density $f_\alpha$ all desired
moments of the distribution function can be calculated, including in particular
the charge density $\rho$ which then allows for the computation of the electric
field.

The discretization can be performed directly for position
$x$ and velocity $v$ in a 1d1v setup or for the components of the respective vectors in higher dimensional implementations of phase space,
as proposed e.g. by \cite{Cheng_1976}. A modern
implementation of this approach is the Vlasiator code (see
\citep{Alfthan_2014}) that contains many clever implementation details to fit
as much of the six dimensional phase space into memory as possible.

The advantage of direct discretization is the very detailed description of
features in phase space, including strongly non-thermal velocity distributions
and the ability to represent both regions with high and low particle density
accurately. The problem is that the phase space density develops structures on
small separations in the velocity direction and consequently large gradients in
$v$. This was first noted by \cite{Lynden_1967} for a similar problem in
galaxy formation but is a general problem with this approach. One possible
solution is periodic smoothing as suggested by \cite{Denavit_1972} and
\cite{Cheng_1976}. However, this can lead to unwanted diffusivity. An
interesting approach to deal with this problem was published in
\cite{Symon_1970}, where a very high resolution in $x$ and $v$ is suggested
to prevent problems with structures at small scales. The trade-off is
that the distribution function at each point in phase space in only represented
by a single bit and density variations are captured using dithering. This
method does not seem to be in use any more, most likely due to the fact that
access to individual bits is relatively expensive on modern CPUs.


An alternative to the direct discretization of position and velocity components
is to transform the phase space density to another suitable basis and
discretize it there. For the spatial direction this is nearly exclusively done
using Fourier transforms, as this makes solving the Poisson equation
relatively easy. For the velocity direction it is possible to use the basis of
Hermite polynomials (see e.g. \citep{Joyce_1971}, \citep{Shoucri_1978} or more
recently \citep{Nakamura_Yabe_1999}) or to use a Fourier transform as well (see
e.g. \citep{Knorr_1963}, \citep{Denavit_1971}, \citep{Klimas_1987}
\citep{Klimas_1994} or the very approachable derivation by \citep{Eliasson_2002}). All these methods trade the problem of the generation of
structures at small scales for the problem of generating modes at high mode
numbers, i.e. high wave numbers in the case of a Fourier transformed velocity direction or alternatively high order Hermite polynomials. The more explicit nature of these features at small scales, however,
allows for a more deliberate treatment.

Beyond the discretization of the phase space density on a grid in $x$-$v$ or
some other suitable orthonormal basis it is also possible to treat the flow of
the incompressible fluid using finite volume methods (see e.g.
\citep{Boris_1976} or \cite{Fijalkow_1999}). The advantage of this method
is that the conservation of phase space density and therefore particle number
can be guaranteed by construction. \cite{Filbet_2003} contains a recent review
on the trade-offs of this approach.

A problem shared by all approaches  that aim to describe the time evolution of the
phase space density as an incompressible fluid is the fact that the six
dimensional phase space is actually fairly high dimensional. Splitting the
amount of available main memory even on a large supercomputer over six
dimensions limits the resolution and dynamic range of each direction. Many
simulation codes therefore neglect either a spatial direction or a velocity direction, dropping the
number of dimensions to five. Removal of a spatial direction requires an assumption about the geometry of the problem, such as the existance of an invariant direction or rotational symmetry. In the case of strong magnetization it is
possible to average over the gyration and remove one velocity component. This
leads to the class of gyrokinetic codes that is widely used in the controlled
fusion community, such as the Gene code (see e.g. \citep{Jenko_2000}). A mathematically careful treatment of the guiding-center model and some other special cases can be found in \cite{Sonnendrücker_1999}.


The main alternative to the Eulerian view of discretizing phase space and
considering the time evolution of the phase space density as the flow of an
incompressible fluid is the (semi-)Lagrangian view of discretizing the phase
space density into independent phase space elements. The challenge is then to
find evolution equations for the phase space elements and to ensure their
correct interaction through the electrcostatic field. In the case where a
constant portion of the phase space density (or weight) is carried by a phase
space element, one ends up with evolution equations that are identical to the
evolution equation for a single, physical particle albeit with modified charge
$q_\alpha$ and mass $m_\alpha$. The ratio between the two quantities is
conserved. This leads to the view of the phase space elements as macro
particles.

The simplest possible simulation code does not even need to solve
for the electric field, but could directly sum the Coulomb force on each
particle that is exerted by all other particles. In the electrostatic limit,
where time retardation effects are absent, this is straight forward. The
downside is of course that the effort to calculate the force on each of the $N$
particles scales as $\mathcal{O}\left(N^2\right)$. A simulation with twice the
number of particles (either due to double the particle density or double the
domain size) will take four times as long. Even on fast computers that are
available today this limits the total simulation size that is feasible.

A possibility to make simulations with a larger number of phase space elements
feasible is to reuse the idea of tree codes that has been employed extensively
in the field of galaxy formation and dynamics (see \citep{Barnes_1986}). As the
name already indicates, particles are stored in a tree-like data structure where
the low order multipole moments of sub-trees are propagated upwards in the tree.
To compute the action on a single particle, the forces from nearby particles
are considered directly (the tree is traversed down to lead nodes) while for
far away sub-trees only the aggregate multipole moments are used. This reduces
the computational effort to  $\mathcal{O}\left(N\,\log(N)\right)$. An
approachable review of this method can be found in \cite{Dubinski_1996}. For
an example on the use in plasma simulations see \cite{Gibbon_2004} and the
description of the highly scalable PEPC code by \cite{Winkel_2012}.

The most widely used option is to deposit the charge density carried by the
phase space elements (on possible other low order moments of the distribution
function such as the electric current) onto a spatial grid and compute the
electrostatic potential on the grid. The Poisson equation can be solved in the
Fourier domain or using by a standard multigrid solver. The
electric field can be obtained from the electric potential by means of numerical differentiation. Alternatively the
electric field can be computed directly from the charge distribution. There is
tradeoff to the exact implementation choice but in any case two important
advantages are realized. The deposition of charge and the calculation of the
force scale as $\mathcal{O}\left(N\right)$ in the particle number $N$ and we
get direct access to the electric field, which allows for additional diagnostics and
physical insights.

The use of Lagrangian phase space samples to compute the time evolution of the
phase space density combined with an Eulerian grid for the computation of the
electric field leads to a wide variety of semi-Lagrangian methods. The main
distinguishing element between the three main families of semi-Lagrangian methods
is the fate of the tracer particles after one time step.

If the phase space elements are kept as is and reused for all further time
steps one ends up with the very popular family of particle-in-cell (PiC) codes.
The big advantage of the PiC method is that the spatial grid never needs to
resolve more than three dimensions and that a rather limited number of
particles per cell suffice to get acceptable coverage of velocity space,
independent of the number of velocity dimensions.
Both the deposition of the charge density $\rho$ onto the grid (by means of one
contribution for every macro particle) and the back reaction of the electric field
onto each particle require interpolation. It
is essential that the same scheme is used for both interpolation steps, to
avoid spurious self forces. The choice of the interpolation, or in other words
the shape of the particle, allows for some trade-off between speed and memory
use on one side and numerical noise on the other side. However, no golden bullet
is available and the limited number of particles per cell leads to
unrealistically large fluctuations in the charge density. Increasing the number
of particles reduces the noise only as $\mathcal{O}\left(N^{-1/2}\right)$. A
very good introduction to PiC codes can be found in the book
\cite{Birdsall_2005} or the in papers that started the field (e.g.
\citep{Dawson_1962}, \citep{Hockney_1966}, \citep{Birdsall_1969},
\citep{Dawson_1983}) or the code descriptions of widely used modern PiC codes
(e.g. \citep{Friedman_1992}, \citep{Fonseca_2002} or \citep{Arber_2015}).
Nearly all PiC codes and a number of other simulation codes use the particle
update scheme defined in the hard to find but excellent paper by
\cite{Boris_1970}.

If keeping the particles leads to a lot of numerical noise, it is of course
tempting to reconstruct the phase space density $f$ from the tracer particles.
From the reconstructed phase space density all desired moments can be
calculated, including the charge density that is used to compute the electric
field for the next update step. It is also used to create new tracer particles
that are advanced for one time step. This method is discussed for example in
\cite{Denavit_1972} and is quite useful to remove beam instabilities that can
occur in plasma simulations with phase space elements at equal spacing
$\Delta{v}$ in the velocity dimension. The downside of the method is that the
reconstruction of $f$ and subsequent resplitting into phase space elements
introduces a lot of numerical diffusion in phase space.

Framed in this way the approach of the Vlasov-Hybrid Simulation method (see
e.g. \citep{Nunn_1990}, \citep{Nunn_1993} or \citep{Kazeminezhad_2003}) becomes
a lot more obvious. In this method, the phase space density is decomposed in
phase space elements that are updated in each time step. From the ensemble of
tracer particles the full phase space density is reconstructed on a grid in
such a way that numerical noise is minimized. From the reconstructed phase space
density $f$ all desired moments can be calculated, including the charge
density. The resulting electric field can be computed following any standard
method and can act on the phase space elements. This approach allows for
schemes with much lower noise, but still uses very simple evolution equations
for the phase space markers instead of solving incompressible fluid equations.
To limit numerical diffusion the same phase space elements are kept and
advected in all further time steps. As this is the method of choice in this
paper a more detailed description is given in the following section.

The fact that such a large number of simulation methods has been developed and
that no single method has won out already shows that the advantages and
disadvantages of the different methods have to be weighted carefully for each
individual problem that is to be studied numerically. Reviews that try to
critically compare the methods go back to at least \cite{Denavit_1971}, but
continue to the present day. This paper is not the place to try to give a final
verdict on the topic. Hopefully the previous paragraphs are however helpful to
give the reader an impression of the multitude of options that are available in
the field of kinetic plasma simulations.

Beyond the kinetic simulations there is the wide and fruitful field of
magnetohydrodynamics (see e.g. \citep{Powell_1999}, \citep{Fryxell_2000}
\citep{Brandenburg_2002} and \citep{Mignone_2007} for papers on popular codes
and references therein as an introduction to the topic). Once fluid description
for plasma species enter the picture it is also possible to consider
multi-fluid descriptions (see e.g.  \citep{Hanasz_2010}, \citep{Glocer_2009} or
\citep{Leake_2012}) or hybrid descriptions that treat some species kinetically
and others as a fluid (see e.g. \citep{Müller_2011}, \citep{Kallio_2002}).

\section{Description of the Method}
\label{sec:method}

\begin{figure}[hbt]
\begin{center}
\includegraphics[width=0.45\textwidth]{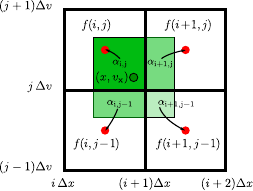}
\end{center}
\caption{Interpolation from a phase space marker at $(x,v_x)$ to the phase space
densities $f$ on the grid follow a standard cloud-in-cell scheme (area
weighting). The particle is described by a cloud of uniform density that is
centered on the particle's position in phase space and that has the same extend
as a grid cell. The overlap with each cell can then be computed easily and
produces the weight $\alpha_\mathrm{ij}$ that determines the contribution of
the particle to the phase space density in the different cells.}
\label{fig:interpolation}
\end{figure}

As mentioned previously the crucial step in this method is the reconstruction
of the phase space density $f$ on a grid in $x-v_\mathrm{x}$ space based on the
Lagrangian phase space markers. The interpolation from a single marker to the
grid cells follow a standard cloud-in-cell (area weighting) scheme that is also
used in many particle-in-cell codes. As illustrated in
Fig.~\ref{fig:interpolation} each particle with position $(x, v_\mathrm{x})$ in
phase space is considered to be a cloud of constant density that spans a $\Delta
x \cdot \Delta v_\mathrm{x}$ area, same as a grid cell. This area overlaps
several grid cells and consequently the particle contributes information about
phase space density to more than one grid point. The relative contributions
$\alpha_\mathrm{ij}$ are calculated by the overlap between the extended particle
and the grid cells. When three velocity components are considered $f$ is
reconstructed on a four dimensional grid in $x-v_\mathrm{x}-v_\mathrm{y}-v_\mathrm{z}$
space and each particle is considered to be a cloud of constant density that
spans a $\Delta x \cdot \Delta v_\mathrm{x} \cdot \Delta v_\mathrm{y} \cdot \Delta v_\mathrm{z}$
volume. The $\Delta v_\mathrm{i}$ are not necessarily equal, but the code does
this by default as it is often convenient to do so. The user is however free to
change this, as it might be useful e.g. in the case of large temperature
anisotropies.

The crucial difference to a particle-in-cell code is the way in which the
contributions of different phase space markers to a single grid point are
combined. Instead of taking a naive sum over marker particles
\begin{equation}
	f\left(i,j\right) = \sum_\mathrm{l} \alpha_\mathrm{ij,l} \cdot f_\mathrm{l} \quad ,
\end{equation}
the Vlasov-Hybrid-Simulation method (VHS) by \cite{Nunn_1993} uses a smarter
reconstruction of $f$:
\begin{equation}
	f\left(i,j\right) = \sum_\mathrm{l} \alpha_\mathrm{ij,l} \cdot f_\mathrm{l} \;\;\;/\;\;\; \sum_\mathrm{l} \alpha_\mathrm{ij,l}\quad .
\label{eqn:reconstruction}
\end{equation}
The difference is rather important. Instead of simply splitting the phase space
density $f_\mathrm{l}$ that is assigned to each particle over a few cells and
depositing it there, the VHS method tries to estimate the true phase space
density $f$ based on the phase space weights assigned to the particle. This
means that particles with low $f_\mathrm{l}$ can actually lower the
reconstructed phase space density $f$ in a cell instead of being drowned out.

This possibility has one important consequence for the initial choice of phase
space markers and the weight they are assigned. In particle-in-cell simulations
usually all markers have the same weight (or macro factor), as particles with
smaller weight cannot effectively modify the deposited charge distribution but
still incur the full computational cost. Phase space regions with low density
will initially be filled with few marker particles and set the limit on the
chosen weight. Regions with high density will be filled with a larger number of
particles that have the same weight and sum to the desired density. The only
trade-off that can be made in the simulation design is between the ability to
represent low density regions and the total particle number or simulation cost.

In a VHS simulation phase space markers can have quite different associated
weights and can represent phase space regions of different densities much more
efficiently. The simple approach chosen for the implementation described below
puts the same number of phase space markers into each grid cell and sets their
weights to produce the desired initial phase space distribution. This is
however not the only possibility and simulations where additional resolution of
some phase space patches later in the simulation is desired might want to track
those patches back to the initial state and add more phase space markers to
these regions.

In a typical one-dimensional particle-in-cell simulation one in 370 particles
will have a velocity outside the range $-3 v_\mathrm{th} \leq v_\mathrm{x} \leq
3 v_\mathrm{th}$. Inside the range there is good coverage with a few hundred
particles per cell, which is easily achievable. A VHS code with typical
parameters of 50 cells per $v_\mathrm{th}$ in $v_\mathrm{x}$ direction and four
particles per phase space cell on the other hand needs 1200 particles per
$\Delta x$ and has the memory overhead from the phase space grid and a slightly
more expensive particle deposition.
However, if coverage out to $\pm5v_\mathrm{th}$ is desired the particle-in-cell
code would need $1.7\cdot10^6$ particles per cell while the number for the VHS
code only increases to a moderate 2000. This nicely demonstrates how much
better the VHS code is at representing low density parts of phase space.
This is even more noticeable when more velocity dimensions are resolved as a
larger fraction of the phase space has small contributions in $f$ as a lot of
grid cells are far away from the mean in at least one velocity dimension.

The ability to capture fine structures and low-density regions in phase space
and the possibility of particles with low weight to effectively moderate the
influence of particles from the center of the distribution with large weights
leads to a scheme that has a very low level of numerical noise in the
reconstructed charge density and resulting electric field. The energy in this
numerical noise of the electric field is much smaller than in a
particle-in-cell simulation. This means that the range of energies between the
kinetic energies of the particles down to the noise floor is much larger.
Consequently the dynamic range over which processes such as Landau damping can
be studied is much larger.

\begin{figure*}[hbt]
\includegraphics[width=\textwidth]{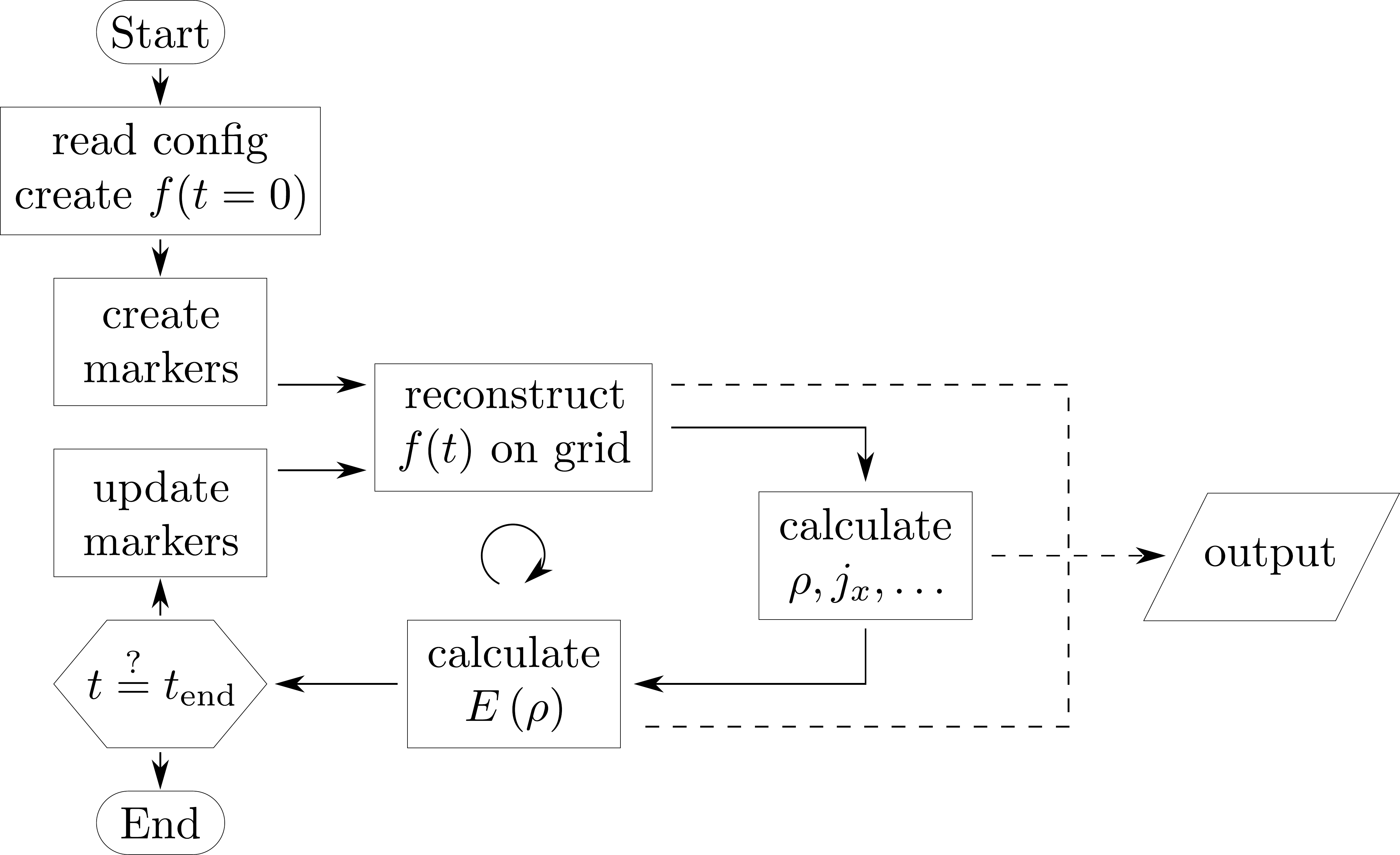}
\caption{
After initialization and marker creation, as well as after every marker update,
the full distribution function $f(x,v_{\mathrm{x}},t)$ is reconstructed on a
grid.
This is the crucial step that allows to calculate moments of the distribution
function such as $\rho(x,t)$ both for output and for the calculation of the
self-consistent, electrostatic field $E(x,t)$.
This field only depends on the net charge density $\rho$ and is calculated
using a spectral solver.
It is then used to update the marker population for the next time-step.
}
\label{fig:algorithm}
\end{figure*}

While the reconstruction of the phase space density that was described so far is
the crucial difference to a particle-in-cell code, the rest of the code follows
the well-known structure of other semi-Lagrangian codes, as shown in
Fig.~\ref{fig:algorithm}.

The code starts by reading the configuration file that describes the physical
properties of the desired initial state, the numerical properties of the
chosen discretization and parameters that describe the desired output. The
exact format is implementation dependent and is described in more detail in
section~\ref{subsec:json}.

The description loaded from the configuration is sufficient to allocate all
necessary data structures and to create the phase space markers that represent
the initial distribution function. Setup of the initial fields is trivial, as
only $E_\mathrm{x}$ is relevant to the particle dynamics. It is set to zero as
no large scale electric fields are expected in a quasi-neutral plasma in a
periodic domain.

Following the initialization the regular time-stepping scheme is started.
First of all the phase space density $f$ is reconstructed on the grid, based on
the Lagrangian marker particles.
Once the reconstructed phase space density is known, it is straight forward to
integrate over the velocity direction(s) of the grid to get moments of the
distribution function, such as charge density $\rho\left(x\right)$ or the
current $j_\mathrm{x}\left(x\right)$. Only the net charge density of all
species together is necessary for the time-stepping, but the individual
contribution of each species is both easier to calculate and can be useful to
study and understand the physics. This is also the reason for calculating higher
moments that don't enter the update cycle directly.
After the net charge density has been obtained it can be used to solve
Poisson's equation for the electrostatic field. Both multi-grid solvers and
spectral solvers work fine. In principle even direct integration should work in
periodic domains.
The resulting electric field is used for output as well as for the calculation
of the force on each marker particle.
For this the electric field on the grid is interpolated to the position of each
particle using the same area weighting scheme that has been used in the
reconstruction step. From the local electric field it is easy to calculate the
force and therefore the acceleration for each particle, which allows to
calculate the new velocity and position. With that step the new state of all
particles is available both for output and to start the whole scheme for an
additional cycle if the desired end of the simulation is not reached yet.
Following this rough overview of the general structure of a code implementing
the VHS method, the next section will go into some of the details and design
choices that went into the implementation that accompanies this paper.

\section{Description of the Implementation}
\label{sec:code}

\subsection{Implementation decisions}
\label{subsec:design}

\begin{figure*}[hbt]
\includegraphics[width=\textwidth]{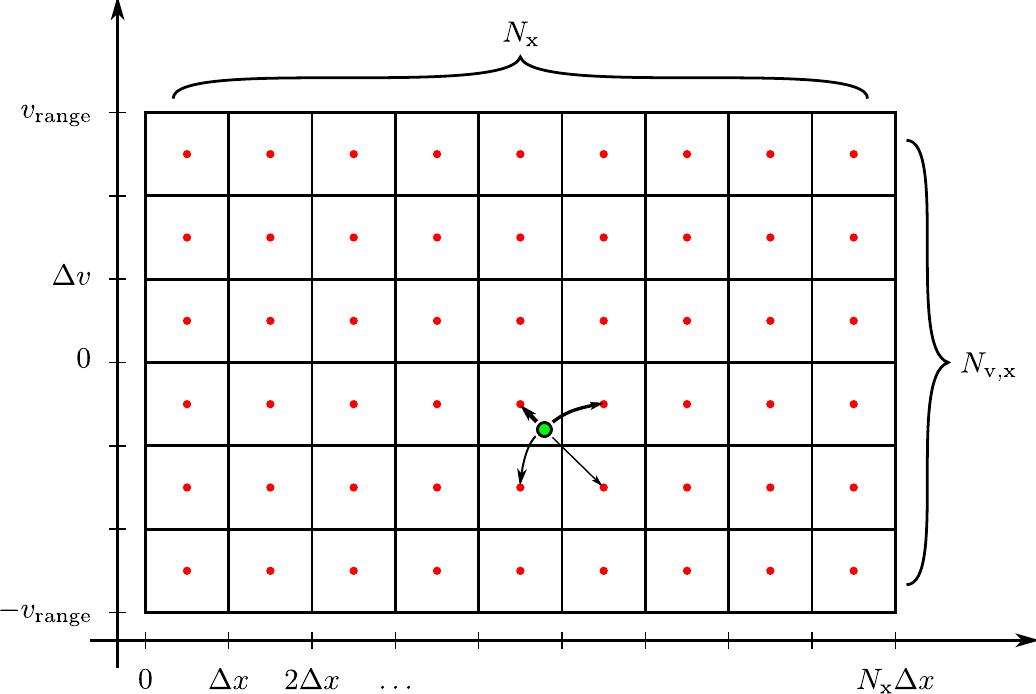}
\caption{
The phase space is discretized in $x$ and $v_{\mathrm{i}}$ directions. In the
spatial direction the domain is split into $N_{\mathrm{x}}$ cells, each $\Delta
x$ wide. In the velocity directions each cell extends over an interval $\Delta v_\mathrm{i}$
and $N_\mathrm{v,i}$ cells cover the range from
$v_{\mathrm{drift}} - 5\,v_{\mathrm{th}}$ to
$v_{\mathrm{drift}} + 5\,v_{\mathrm{th}}$. The initial drift and thermal speeds can be different in different velocity components, allowing for the study of e.g. temperature anisotropies.
All grid quantities are collocated
at the center of the grid cells, as indicated by the red dots. Particle based
quantities (indicated by the dot with green fill and black outline) are
deposited onto the grid using the cloud-in-cell interpolation method (area
weighting) described in Fig.~\ref{fig:interpolation}.
}
\label{fig:phasespace}
\end{figure*}

A lot of operations in the code require interaction with the reconstructed
distribution function on the phase space grid. An important point in the design
is therefore the choice of the discretization of phase space.
Figure~\ref{fig:phasespace} shows what the accompanying code uses. The spatial
dimension is split into $N_\mathrm{x}$ cells that span the range from $0$ to
$L$. Each cell has a width of $\Delta x = L / N_\mathrm{x}$. Each resolved velocity
direction is split into $N_\mathrm{v,i}$ cells and spans from $v_{\mathrm{drift,i}}
- 5\,v_{\mathrm{th,i}}$ to $v_{\mathrm{drift,i}} + 5\,v_{\mathrm{th,i}}$. Each
species has its own velocity range, covering ten times the thermal speed
$v_\mathrm{vth,i}$, centered around the mean drift speed $v_{\mathrm{drift,i}}$.
Different species thus use different values of $\Delta v_\mathrm{i}$ but the number of
cells in velocity direction is the same for all species. Every gridded quantity
is represented by the single value located at the center of the grid cell.
As mentioned earlier and shown in Figure~\ref{fig:interpolation}, area
weighting is used to interpolate between the Eulerian view and the Lagrangian
marker particles.

Gridded quantities, such as momenta of the distribution function use a grid
that is the same $N_\mathrm{x}$ cells of width $\Delta x$ and are represented
by the value in the center of the grid cell. This way they can be calculated
through summation over the velocity directions without further
interpolation. The electric field uses the same grid layout. This collocates
the electric field values with the net charge density, which is convenient for
the spectral solver. Furthermore it makes the interpolation going back to the
particles identical to the interpolation in the reconstruction step, which
avoids confusion.

As already mentioned, the code uses a spectral solver to calculate
$E_\mathrm{x}\left(x\right)$ based on $\rho\left(x\right)$. It relies on the
fact that the differential equation for the electric field given by Eq.
\eqref{eqn:poisson} reduces to an algebraic equation in $k$ space:
\begin{equation}
\tilde E\left(k_\mathrm{x}\right) = - \frac{4 \pi \, i \, k_\mathrm{x}}{k^2} \tilde \rho\left(k_\mathrm{x}\right) \quad .
\label{eqn:poissonk}
\end{equation}
In the implementation it takes one Fourier transform to go from the charge
density $\rho$ to $\tilde \rho$. The electric field $\tilde E$ as a function of
$k$ can then be easily obtained following Eq. \eqref{eqn:poissonk}. The
resulting quantity is transformed back by a second Fourier transform
to calculate $E$. The implementation in the code contains two little twists.
One is the removal of the DC component by setting $\tilde
E\left(k_\mathrm{x}=0\right)$ to zero. This removes the division by zero for
$k_\mathrm{x}=0$ and is physically well motivated as no large scale fields are
expected, especially for a plasma without net charge in a periodic domain. The
second is a modification of the differencing operator in $k$-space:
\begin{subequations}
\label{eqn:kappasoftening}
\begin{align}
\tilde E\left(k_\mathrm{x}\right) &= - \frac{4 \pi \, i \, \kappa\left(k_\mathrm{x}\right)}{K^2\left(k_\mathrm{x}\right)} \tilde \rho\left(k_\mathrm{x}\right) \quad \quad \quad, \, \\
\kappa\left(k_\mathrm{x}\right) &= k_\mathrm{x} \,\cdot \, \mathrm{sinc}{\left(k_\mathrm{x}\cdot\Delta x\right)}\quad \quad \quad \,\,, \\
K^2\left(k_\mathrm{x}\right) &= k^2_\mathrm{x} \,\cdot \, \mathrm{sinc}^2{\left(0.5 \cdot k_\mathrm{x}\cdot\Delta x\right)} \quad .
\end{align}
\end{subequations}

This modification is sometimes called ``sinc softening'' and makes the
differencing operator act more like a local operator. This fits in well with
the finite differences that are used in the rest of the code. A nice overview
of the trade-offs between Eq. \eqref{eqn:poissonk} and
\eqref{eqn:kappasoftening} can be found in \cite[Appendix E]{Birdsall_2005}.

As indicated in Fig.~\ref{fig:algorithm} the code doesn't directly create the
marker particles based on the description in the configuration file, but
calculates the initial distribution function on the phase space grid. Following
this step, Lagrangian phase space markers are created to represent this
distribution. The intermediate step of discretizing the distribution function
allows to separate the description of the initial state from the creation of
the marker particles in a clean way, which makes the extension to other initial
states easy.


As indicated in Fig.~\ref{fig:phasespace} the phase space is discretized in
steps of $\Delta v$ in the velocity direction. As mentioned earlier, the Vlasov
equation leads to the creation of finer structures in velocity space over time.
When these structures reach the limited resolution of the numerical simulation
the validity of the simulation result becomes questionable. This widly known
recurrence problem (see e.g. \citep{Canosa_1972}, \citep{Canosa_1974} or
\citep{Manfredi_1997}) is not solved in any innovative way in this code.
However the recurrence time $T_R$ until the problem appears is given by
\begin{equation}
	T_R = \frac{\pi}{k\,\Delta{v}} \quad ,
	\label{eqn:def_recurrance_time}
\end{equation}
where $k$ is the wave number of spatial variations and $\Delta v$ the
resolution in the velocity direction. The smallest wave number is set by the
spatial resolution $\Delta x$ multiplied by the number of spatial grid points $N_x$. The
velocity spacing is automatically calculated by the code from the thermal
speed $v_\mathrm{th}$ and the number of grid points $N_\mathrm{v,x}$. Inserting this into
Eq.~\eqref{eqn:def_recurrance_time} gives the recurrence time
\begin{equation}
	T_R = \frac{1}{10} \frac{\Delta x}{v_\mathrm{th,x}} N_\mathrm{x} N_\mathrm{v,x} \quad ,
	\label{eqn:recurrance_time_code}
\end{equation}
i.e. the time at which the initial distribution function reoccurs.
The code calculates and prints this value and gives a strong warning if a
simulation with a total simulated time $T = N_t\,\Delta{t} > T_R$ is attempted. As the
VHS model is mostly used if a lot of
resolution in velocity space is desired, $N_\mathrm{v,x}$ will tend to be large
and the recurrence effect can be expected to happen very late, past the desired end time of the
simulation. If longer times need to be simulated and one is
not interested in features on small scales in the velocity direction, the
addition of a carefully chosen level of collisionality following the approach
by \cite{Pezzi_2016} could help.


For all the test cases presented in section~\ref{sec:tests} the
recurrence time is between 6 times (for the long run on probing Landau damping
in subsection~\ref{subsec:landau}) to 1399 times (in the test case determining
the dispersion relation of Langmuir waves in subsection~\ref{subsec:dispersion})
longer than the total duration of the simulation. Recurrence effects are
therefore not expected to affect any of the test results.


\subsection{Performance and Parallelization}
\label{subsec:perf}

The performance of the code is dominated by the time spent on depositing marker
particles onto the phase space grid and updating the particles based on the
local electric field. Given that the number of particles per cell doesn't
change much -- it is always between 2 and 20 -- the effort to compute the next
time step is directly related to the resolution of the phase space grid. To
compare with other semi-Lagrangian codes it is however convenient to specify
the performance through the number of particles that are handled per second and
CPU core. On an Intel Xeon E5-2650 CPU running at 2.6 GHz the code is capable
of about $10^7$ particle updates per second using a single CPU core. This is
slightly slower than expected for a particle-in-cell code, but not drastically
so.

The majority of the time -- slightly more than 2/3 -- is spent on the
reconstruction of the phase space distribution. This is comparable to a
particle-in-cell code, where the deposition of the charge density is a major cost.
Calculating the updated velocity and position of the particles is the second
biggest cost at about 1/5 of the CPU time. Inclusion of additional velocity
components increases this fraction, especially as the action of static
background magnetic fields requires the implementation of a full Boris push to update the particles.
The cost of the spectral solver that calculates the electric field is
negligible even for larger simulations despite its
$\mathcal{O}\left(n\,\log\left(n\right)\right)$ scaling. Even for simulations
with low grid resolution in the $v_\mathrm{x}$ direction and few marker
particles per cell it requires well below five percent of the CPU time.
The remaining 10 to 15 percent of the CPU time are used by the calculation of
moments of the distribution function which requires integration over the
reconstructed distribution function as well as output and internal book
keeping.

\begin{figure}[htb]
\begin{center}
\includegraphics[width=\columnwidth]{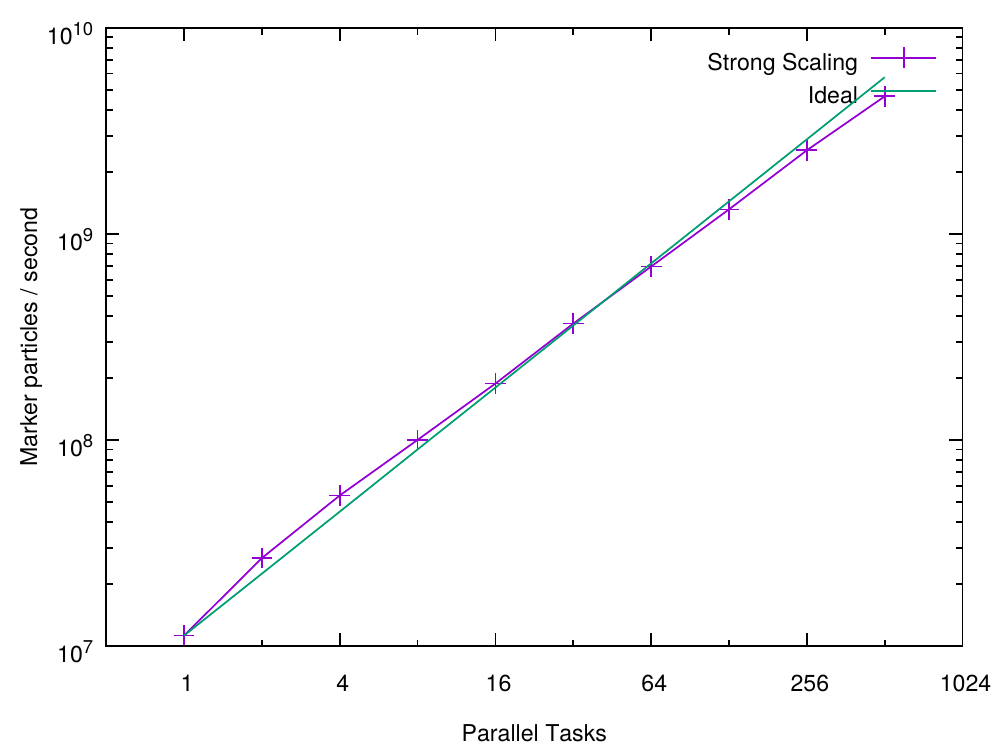}
\end{center}
\caption{
The strong scaling test simulates a phase space grid with 8192 cells in $x$
direction and 8192 in $v_\mathrm{x}$ direction using electrons and protons
represented by nine particles per cell each. The simulation is performed using
a variable number of MPI threads and the average performance over 1500 time
steps is given, excluding output.
}
\label{fig:strong_scaling}
\end{figure}

Figure~\ref{fig:strong_scaling} shows the performance of the code when
running the same simulation with a variable number of parallel tasks. The
simulation discretizes phase space with $2^{13}$ cells in both the $x$ and
the $v_\mathrm{x}$ direction. Electrons and protons in the simulation are
represented by nine marker particles per cell which leads to a grand total of $1.2
\cdot 10^9$ particles in the simulation that are updated for 1500 time steps.
Given that the field solver incurs negligible computational cost the code
should exhibit a linear performance increase when adding additional CPU cores.
The code shows good scaling efficiency to at least 256 CPUs.

\begin{figure}[hbt]
\begin{center}
\includegraphics[width=\columnwidth]{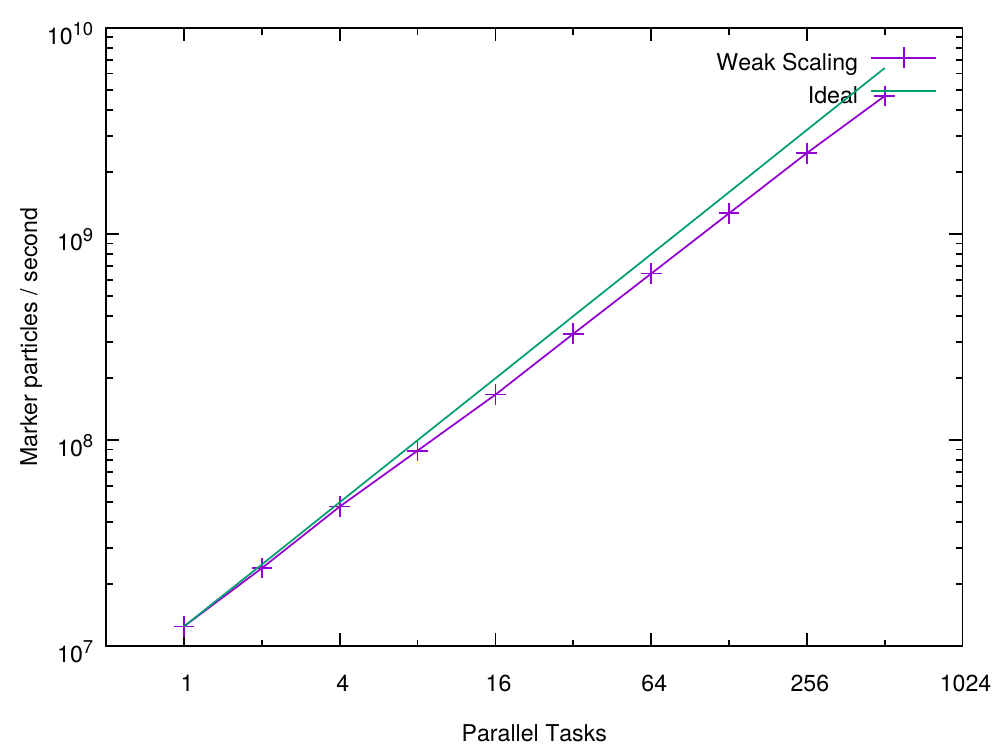}
\end{center}
\caption{
The weak scaling test also simulates a phase space with 8192 cells in
$v_\mathrm{x}$ direction and nine particles per cell electrons
and protons each. The simulation size in $x$ direction however is changed with the
number of MPI threads so that each parallel task is responsible for 16 cells
in $x$ direction. Performance is again measured by taking the average over a
duration of 1500 time steps without performing output.
}
\label{fig:weak_scaling}
\end{figure}

In many cases the goal is not shortest time to solution but to simulate the
largest domain that is feasible. This is especially true for plasma simulation
codes that have to resolve the micro-physical length scale set by the Debye
length and struggle to reach the larger outer length scales. In this case weak
scaling, where the amount of work per parallel thread is held constant, is much
more interesting. Figure~\ref{fig:weak_scaling} shows the performance of the
code in the case where additional CPUs are added to allow additional parallel
tasks to simulate a larger and larger simulation domain. Scaling efficiency is
again very good in this case. The code
is able to make efficient use of typical clusters available to researchers at
their department if the problem size is reasonable.

\subsection{Format of the configuration files}
\label{subsec:json}

The configuration files use JavaScript Object Notation (JSON). This format is
easy to read and write for users of the code, without requiring (much)
experience with programming. Nevertheless it is easy to parse or generate
automatically and a wide variety of libraries that can do so are available for
many different programming languages. This allows for easy use of the
configuration file by other programs. Both graphical front ends that generate
configuration files and programs that operate on the simulation output, such as
automated analysis scripts, can handle the file format easily.

The configuration file is split into three parts titled ``physical'',
``numerical'' and ``output''. Examples for each part are given in
listings~\ref{code:physical} to \ref{code:output}. Note that the order of parameters and
sections in the JSON file doesn't change how the file is parsed. Proper nesting
of blocks on the other hand is crucial as the file won't form valid JSON and
cannot be parsed otherwise.

\begin{lstlisting}[language=json,label=code:physical,caption=Section of a configuration file describing the physical setup,lastline=17,name=foo]
{
  "physical": {
    "species": {
      "electrons": {
        "q":      -4.8032e-10,
        "m":       9.1094e-28,
        "omegap":  1.0000e9,
        "vth":     1.4990e9,
        "vdriftx": 0.0,
        "knum":    1.0,
        "alpha":   0.05
      }
    },
    "Bx": 0.0,
    "By": 0.0,
    "Bz": 0.0
  },
\end{lstlisting}

The first section deals with a description of the physical plasma conditions.
All parameters in this section should make sense irrespective of the simulation
method that is used. The parameter values are given in CGS units.
In the accompanying code, the object that describes the physical setup only
contains an object describing the different species in the plasma. In versions
of the code that resolve more than one velocity component the background
magnetic field is also specified in this section.

Each species object contains key-value pairs, where the key is a string that
gives a name to the species and the value is a JSON object containing all
parameters describing the species. The species name is also used to name the
output about this species and therefore has to be a valid name for a HDF5
group. In practice one should limit the name to  alphanumerical ASCII
characters supplemented by underscore, dash and plus. Other characters (such as
square brackets) might be allowed under the standard but can lead to problems.

The parameter \texttt{q} gives the charge $q_\alpha$ for each physical particle
of this species in statcoulomb. A marker particle of this species in the code
might have a much larger charge, but this is transparent to the user.

The parameter \texttt{m} works in the same way and gives the mass $m_\alpha$ of
a physical particle in grams. The ratio $q_\alpha/m_\alpha$ is constant for all
marker particles of a species, independent of phase space weight.

The parameter \texttt{omegap} gives the plasma frequency
$\omega_\mathrm{p,\alpha}$ for species $\alpha$ in \SI{}{\radian\per\second}.
This implicitly sets the density of this species through
\begin{equation}
n_\alpha = \left(\omega^2_\mathrm{p,\alpha} \, m_\alpha\right) / \left(4 \pi \, q^2_\alpha\right) \quad .
\end{equation}
The overall plasma frequency $\omega_\mathrm{p}$ is often useful and is given
by the root of the sum of the squares of the individual plasma frequencies of
all species:
\begin{equation}
\omega_\mathrm{p} = \sqrt{\sum_\alpha \omega^2_\mathrm{p,\alpha}} \quad.
\end{equation}

The fourth and last parameter that is necessary for every species is
\texttt{vth} that sets the thermal speed $v_\mathrm{th,\alpha}$ in centimeters
per second. This is used as standard deviation $\sigma$ in the normal
distribution that describes the initial velocity distribution. It is common,
but not necessary for the code, that different species are in thermal
equilibrium. In this case the thermal speeds should scale as
$v_\mathrm{th,\alpha} \propto m_\alpha^{-1/2}$. This parameter also sets the
limits of the phase space grid for the species. The maximum and minimum
velocities $v^{\pm}_\mathrm{range}$ (see Fig.~\ref{fig:phasespace}) are set to
$v_\mathrm{drift,\alpha}\pm 5 \cdot v_\mathrm{th,\alpha}$. The thermal speed of
the lightest species and the overall plasma frequency $\omega_\mathrm{p}$ set
the Debye length $\lambda_\mathrm{D} = v_\mathrm{th} / \omega_\mathrm{p}$,
which is the natural length scale in the plasma.
If the 1d3v version of the code is used and a temperature anisotropy is
desired, it is possible to set \texttt{vthx}, \texttt{vthy} and \texttt{vthz}
separately instead of using \texttt{vth} which will set the thermal speed in
all direction to the same value.

The mean drift speed of each species that was just mentioned is an optional
parameter that defaults to zero, i.e. a non drifting population. Using the key
\texttt{vdriftx} (or \texttt{vdrifty} or \texttt{vdriftz} in the 1d3v version of the code)
it can, however, be set to a value in centimeters per second and
will then set the average speed of that species. This allows for the
initialization of beams, counter propagating neutral plasmas, or net currents.

The two remaining parameters that are shown in the example are also optional.
If the parameter \texttt{alpha} is given and non-zero, a sinusoidal perturbation
is added to the initial density distribution:
\begin{equation}
n_\alpha\left(x\right) = \frac{\omega^2_\mathrm{p,\alpha} \, m_\alpha}{4 \pi \, q^2_\alpha} \cdot \left(1 + \mathtt{alpha} \cdot \sin{\left(\frac{2 \pi \, \mathtt{knum} \, x}{L} \right)}\right) \quad .
\end{equation}
The parameter \texttt{knum} determines how many wavelengths of the perturbation
fit into the simulation box of length $L$. 

\vspace{\baselineskip}

The second section contains parameters relevant to the numerical discretization
of the problem, such as the number of grid cells. All parameters in this
section are dimensionless.

\begin{lstlisting}[language=json,label=code:numerical,caption=Section of a configuration file describing the numerical parameters chosen for the discretization,name=foo,firstnumber=last,firstline=18,lastline=25]
  "numerical": {
    "rescale_dx" : 1.01,
    "rescale_dt" : 1.01,
    "Nx" :         1024,
    "Nv" :         2048,
    "Nt" :         1500,
    "ppc" :           8
  },
\end{lstlisting}

The code is able to make a good guess what a reasonable cell size $\Delta x$
should be based on the physical parameters, but it can be modified through the
optional parameter \texttt{rescale\_dx}. Values greater than one lead to smaller
grid cells and better resolution:
$\Delta x = 0.4 \, \lambda_\mathrm{D} / \mathtt{rescale\_dx}$.

A similar parameter \texttt{rescale\_dt} exists to adjust the length of the
time step $\Delta t$. It is mostly based on the plasma frequency and is given
by $\Delta t = 0.25 \pi \, / \, \omega_\mathrm{p} \, / \,
\mathtt{rescale\_dt}$. In the magnetized case the time step is chosen small
enough to resolve the particles' gyrations in the prescribed magnetic field with at least $2\,\pi$ steps per gyration.

With the size of a grid cell given by $\Delta x$ it is still necessary to
specify the number of such cells $N_\mathrm{x}$. This is done by the parameter
\texttt{Nx} in the configuration file. The total length of the simulated domain
is given by $L = N_\mathrm{x} \cdot \Delta x$.

Similarly the number of grid cells in the velocity direction is specified by
\texttt{Nv}. Note that the number of grid cells is the same for every species
but the grid spacing is given by
$\Delta v_\mathrm{x,\alpha} = 10 \, v_\mathrm{th,\alpha} \, / \, N_\mathrm{v}$
and can be different for each species. If desired it is possible to set the
number of grid cells in each velocity direction separately by using
\texttt{Nvx}, \texttt{Nvy} and \texttt{Nvz}. By default the code uses the
value supplied in \texttt{Nv} for each resolved velocity dimension.

The length of each time step $\Delta t$ is mostly fixed by physical parameters.
The duration of the simulation $T$ is therefore mostly determined by the number
of time steps $N_\mathrm{t}$ that is given by the parameter \texttt{Nt}.

The last remaining numerical parameter is \texttt{ppc}. It controls the number
of marker particles for each species that is initialized per $\Delta x \cdot
\Delta v_\mathrm{x}$ cell in phase space. This number can be much smaller than
in a particle-in-cell code, but keep in mind that the total number of particles
$N_\mathrm{p}$ is given by $N_\mathrm{x} \cdot \prod_i N_\mathrm{v_i} \cdot
\mathtt{ppc}$ which includes the product of the number of cells in all resolved velocity
directions. This extra factor is missing in a particle-in-cell code and can be large, especially in the case of the 1d3v code.

\vspace{\baselineskip}

The last section of the configuration file controls output settings. The gap
between computing speed and storage speed has continuously broadened in
computers over the last decades. At this point the code would be completely
limited by the write speed of the I/O subsystem without using a significant
fraction of the available processing power if it tried to store every available
diagnostic in every single time step. Limiting the output to the desired
quantities at reasonable cadence is much more efficient.

\begin{lstlisting}[language=json,label=code:output,caption=Section of a configuration file describing the desired output quantities,name=foo,firstnumber=last,firstline=26]
  "output": {
    "dir":     "/scratch/vhs/",
    "output_every":              0,
    "energyout_every":          10,
    "fields_every":              1,
    "fields_per_species_every":100,
    "phasespace_every":        500
  }
}
\end{lstlisting}

The parameter \texttt{dir} controls to which directory the output is written.
As the code is using parallel HDF5 the directory needs to be on a file system
that is reachable by all nodes in a parallel run and for which MPI-IO is
possible.

The parameter \texttt{output\_every} controls the cadence of all output types
unless a different cadence is given for the specific output. The parameter is
given as the number of time steps between consecutive outputs. If the value is
zero, the output is disabled.

The code does offer output of integrated thermodynamical quantities (see
\ref{subsubsec:energy_output}) as ASCII data. If the parameter
\texttt{energyout\_every} is given it overwrites the cadence given by
\texttt{output\_every}.

The code also offers output of spatially resolved data, such as the charge
density $\rho$, the current density $j_\mathrm{x}$, or the electric field
$E_\mathrm{x}$ as HDF5 datasets. The cadence of such output is controlled by
the parameter \texttt{fields\_every}.

In addition to the fields that concern all species, such as the net charge
density $\rho$ or the electric field $E_\mathrm{x}$, the code can also output
quantities that are related to a single species such as the charge density for
that species $\rho_\alpha$ or the current carried by it $j_\mathrm{x,\alpha}$.
These quantities do not directly affect the time evolution, but can be very
useful to distinguish physical processes. The frequency of this kind of output
is controlled by the parameter \texttt{fields\_per\_species\_every}.

The most verbose kind of output that is possible is the output of snapshots of
the reconstructed distribution function of the two (or even four) dimensional phase space
grid. This output is very useful to diagnose and discuss phenomena such as
phase space holes. The downside is of course the much larger output size.
Therefore this kind of output should not be requested too often. The output is
done for every species separately and can be controlled by the parameter
\texttt{phasespace\_every}.

\subsection{Output format}
\label{subsec:output}

As already mentioned in the discussion of parameters in the configuration file
the code supports two different kinds of output. The first is the output of
thermodynamical quantities that are integrated over the whole domain. Due to the
small size and to allow access to this information while the code is running
this output is written in ASCII format as one line per output time step. The
format of the line and the meaning of the different columns is described in
\ref{subsubsec:energy_output}.

The other kind of output involves data that is resolved in the spatial and
possibly velocity direction. Due to the larger output size and for
compatibility with the ACRONYM code (see \cite{Kilian_2012}) this is written in
the form of datasets inside a HDF5 file. The file acts as a container that
contains a group for every time step that produces output. Inside the group all
output for this time step is located. More details on the different kinds of output
contained in the HDF5 files follow in \ref{subsubsec:fields_output} and
\ref{subsubsec:phasespace_output}.

\subsubsection{Thermodynamical quantities}
\label{subsubsec:energy_output}

The ASCII output of thermodynamical quantities is stored in a file called
\texttt{energyout.dat} in the output directory. It starts with a header that
describes the format of the file:

\begin{lstlisting}
#step t Ee Ekin Etotal Ekin_1... f_1... S_1...
\end{lstlisting}

The first column contains the integer time step \texttt{step} in which the
output was performed.
The second column contains the same information in form of the time $t$ that
has passed since the beginning of the simulation. Physical time is normalized to
the plasma frequency and is given by
$t = \mathtt{step} \cdot \Delta t \cdot \omega_\mathrm{p}$.
As such it is much better suited to plot or discuss physical processes but it
is not as handy for discussions about the code.

The third column contains the energy that is contained in the electric field in
\SI{}{\erg} which is calculated from
\begin{equation}
E_\mathrm{e} = \frac{1}{8 \pi} \int E^2_\mathrm{x} \, \mathrm{d}V = \frac{1}{8 \pi} \sum_{i=0}^{N_\mathrm{x}} E^2_\mathrm{x} \, \mathrm{d}x^3 \quad .
\end{equation}

The fourth column contains the total kinetic energy of all particles in the
simulation. It is calculated in the non-relativistic approximation as particles
moving close to the speed of light would produce sufficient currents and
magnetic fields that the electrostatic approximation is not applicable any
longer. The output is given as energy in \SI{}{\erg}.

The fifth column contains the total energy in the simulation box. It is
calculated as the sum of electric and kinetic energy. Given that a closed
system in a domain with periodic boundaries is simulated this quantity should
stay constant throughout the simulation run. Small deviations due to the
discretization of phase space are possible. Larger deviations indicate a lack
of resolution.

After the previous five columns that always exist in that order independent of
the number of different species in the simulation follow the kinetic energies
of the different species. The total kinetic energy in column four is actually
the sum of all these. Having this value broken down into one column per species
can be useful to study the energy transfer between species.

In the next columns the integral over $f$ for each species is given. This quantity
is equivalent to the total number of particles of that species:
$N_\alpha\left(t\right) = \int_0^L \int_{-v_\mathrm{range}}^{v_\mathrm{range}} f\left(x, v_\mathrm{x}, t\right) \, \mathrm{d}v_\mathrm{x} \, \mathrm{d}x$.
This quantity should of course be conserved. Only if the time evolution
produces fine structures in phase space that cannot be represented on the grid
during the reconstruction of the phase space density deviations are noticeable.

The last set of columns is the (coarse-grained) entropy per species, calculated
by
\begin{equation}
	S_\alpha = \int_0^L \int_{-v_\mathrm{range}}^{v_\mathrm{range}} f\left(x, v_\mathrm{x}, t\right) \, \log{\left(f\left(x, v_\mathrm{x}, t\right)\right)} \, \mathrm{d}v_\mathrm{x} \, \mathrm{d}x \quad .
\end{equation}
In a completely collisionless plasma entropy should be conserved. Due to the
finite grid resolution there is some coarse graining which allows changes in
$S_\alpha$. However if the changes are larger than a few percent over the
course of the simulation it is an indication that the resolution in phase space
is not sufficient.

\subsubsection{Spatial output}
\label{subsubsec:fields_output}

Beyond the spatially integrated quantities discussed in
section~\ref{subsubsec:energy_output} the code also outputs spatially resolved
quantities. The larger volume of these make collection on a single CPU as well
as output in ASCII format rather inefficient. Therefore parallel output to a
single file is performed in HDF5 format using MPI-IO. The output of all time
steps is stored in that file, grouped by the time step it belongs to.

Among the things that are stored is the net charge density
\begin{equation}
\rho\left(x\right) = \sum_\alpha q_\alpha \int_{-v_\mathrm{range}}^{v_\mathrm{range}} f_\alpha\left(x,v_\mathrm{v}\right) \, \mathrm{d}v_\mathrm{x} \quad .
\end{equation}
It is given in units of \SI{}{\statcoulomb\per\centi\meter\cubed}. The
resulting dataset is named \texttt{rhoL[0]}.

Although not used for the time evolution the current density
\begin{equation}
j_\mathrm{x}\left(x\right) = \sum_\alpha q_\alpha \int_{-v_\mathrm{range}}^{v_\mathrm{range}} v_\mathrm{v} \cdot f_\alpha\left(x,v_\mathrm{v}\right) \, \mathrm{d}v_\mathrm{x}
\end{equation}
is also computed and added to the output. It is given in units of
\SI{}{\statcoulomb\per\centi\meter\squared\per\second} and stored under the
name \texttt{rhoL[1]}. The 1d3v code also represents the velocity components
in $y$ and $z$ direction and stores the corresponding current components in
\texttt{rhoL[2]} and \texttt{rhoL[3]}.

The next quantity that is represented in the output is the electric field
$E_\mathrm{x}$ that is calculated using the spectral solver. It is given in
units of \SI{}{\statvolt\per\centi\meter}. For fields the numbering is slightly
different than for densities and the field is sorted under the name
\texttt{E[0]}.

The last output that is stored as field output is the remaining error of the
field solver. It is given as the residual charge $\Delta \rho = \rho -
\frac{\partial}{\partial x} E_\mathrm{x} / \left(4 \pi\right)$ that is not
accounted for in the electric field and has the same units as a charge density.
The resulting dataset is named \texttt{diff\_rho\_divE0}. This quantity is
mostly useful when experimenting with the field solver. Note that deviations
from zero might not only arise from the field solver but also from the finite
differences approximation for the divergence of the electric field.

If requested the output of spatially resolved quantities will also contain
quantities about individual species $\alpha$. This includes the charge density
$\rho_\alpha$ contributed by this species as well as the current density
$j_\mathrm{x,\alpha}$ carried by it. The datasets are in the same units and have
the same name as their counterparts that are summed over all species. The path
to the datasets in the HDF5 file however contains the name of the species to
allow correct attribution.

One additional dataset is available in the output, which  is broken down by
species, but which is not available as a net quantity. This is the second
moment of the distribution function that can be viewed as a measure for the
temperature or mean squared velocity. It is given by $q_\alpha \, n_\alpha \,
v^2_\mathrm{th,\alpha}$ and stored in units of
\SI{}{\statcoulomb\per\centi\meter\per\second\squared} under the name of
\texttt{rhoL[4]}. Note that in the 1d3v code the trace of the second order
tensor is given which is equivalent to the average temperature. Users who are
interested in the study of temperature anisotropies might want to split out
the average squared velocities for each of the three components or even
reconstruct all six independent components of the second order tensor and
output them separately.

\subsubsection{Phase space output}
\label{subsubsec:phasespace_output}

The most detailed output that is possible is that of quantities that are
reconstructed on the phase space grid. This is always computed and stored
separately for each species. In principle all other output can be calculated
from the full resolved data, but doing so would increase the output size and
post-processing time by a large factor. At lower cadence however this type of
output can provide useful insights.

The obvious quantity to consider on the phase space grid is the reconstructed
phase space density $f_\alpha$. This is available for each species under the
name \texttt{f} in units of \SI{}{\per\centi\meter\cubed}.
Additionally available is \texttt{n} that is given by the denominator in
Eq. \eqref{eqn:reconstruction}. It gives the sum of interpolation weights
of all marker particles that were considered in the reconstruction. It can be
considered as the area weighted form of the number of marker particles that
contributed. Grid cells that show values much below the value configured
through \texttt{ppc} suffer from poor representation and the reconstruction
might be poor there.

The third quantity that is available is the change of $f$ compared to the
initial distribution function $f_0$ that was generated based on the description
in the configuration file. It is calculated as $\delta f = f - f_0$ and stored
under the name \texttt{df}. This quantity is provided mainly as a convenience
feature as plots where the main Maxwellian has been removed are much easier to
interpret.

As different species might cover different velocity ranges that are not
immediately obvious from the numerical grid in phase space, the group containing
\texttt{f}, \texttt{df} and \texttt{n} also contains the upper and lower
limits of velocity space that is covered by these quantities. The two values per dimension
are named \texttt{vmin\_i} and \texttt{vmax\_i} and give the minimum velocity at the
lower end of the first cell and the maximum velocity at the upper end of the
last cell for velocity direction $i$ in units of \SI{}{\centi\meter\per\second}.

\section{Results from the Test Cases}
\label{sec:tests}

The number of scenarios that can be simulated with a one-dimensional
electrostatic code in periodic boundaries is not very large. This is not a
severe problem as the accompanying code is mostly meant as a starting point
into a not very widely used simulation method. It does however make testing of
the code difficult. As the simplest test we consider -- in subsection~\ref{subsec:dispersion} -- the self-consistent
evolution of Langmuir waves that evolve out of mall random initial perturbations.
As a second test
subsection~\ref{subsec:landau} considers the Landau damping of a single
Langmuir wave that is driven by an initial perturbation in charge density. To
test effects of drifting species the current-driven ion-acoustic instability is
shown in section~\ref{subsec:cdiai}. To demonstrate and test the 1d3v
capabilities of the code section~\ref{subsec:bernstein} shows the dispersion
relation of electron Bernstein modes which require the gyration of electrons in
a uniform magnetic field oriented perpendicular to the resolved spatial
direction. The setup file for all test cases are supplied alongside with the
code.

\subsection{Dispersion relation of Langmuir waves}
\label{subsec:dispersion}

A test that is very simple from a physical point of view, but that tests large
parts of the code and their interplay, is a simple neutral and homogeneous
plasma with two species in thermal equilibrium. In the continuum case of Vlasov
theory this will not produce any electromagnetic fields. A real physical
system as well as a numerical system with discrete charges however will produce
some fluctuations in the electromagnetic fields (see \cite{Sitenko_1967}).
These fluctuations are sharply concentrated on the linear eigenmodes of the
system. Only two eigenmodes are accessible to the system under study. One are
Langmuir waves that can be seen as the warm plasma extension of plasma
oscillations. The other possibility are ion acoustic waves. These however
happen at lower frequencies and consequently require longer simulations. This
first test therefore concentrates on Langmuir waves to save computational
effort.

While the test is very simple on the physics side it needs a lot of the code to
be correct and to interact correctly. Wrong normalization in the
reconstruction of the phase space density or in the field solver lead to a
wrong plasma frequency which is easy to spot in the simulation output. Wrong
interpolation leads to self forces on the marker particles which manifest as
parallel stripes in the dispersion plot. Wrong normalizations of space or time
steps lead to a wrong velocity scale which can be seen as a deviation of the
wave's velocity in intermediate $k$ ranges. It takes some experience in code
building to guess the location of a bug from the characteristic deviations in
the dispersion relation, but if the code passes the test it is a good
reassurance that there are no major issues with the code.

One additional very good feature of this test case is that no elaborate
diagnostics are necessary. Taking the spatially resolved output of the
electrostatic field at different time steps and performing Fourier
transformations over space and time readily gives the distribution of the
energy density in these fluctuations as a function of $k_\mathrm{x}$ and
$\omega$.

\begin{figure*}[hbt]
\includegraphics[width=\textwidth]{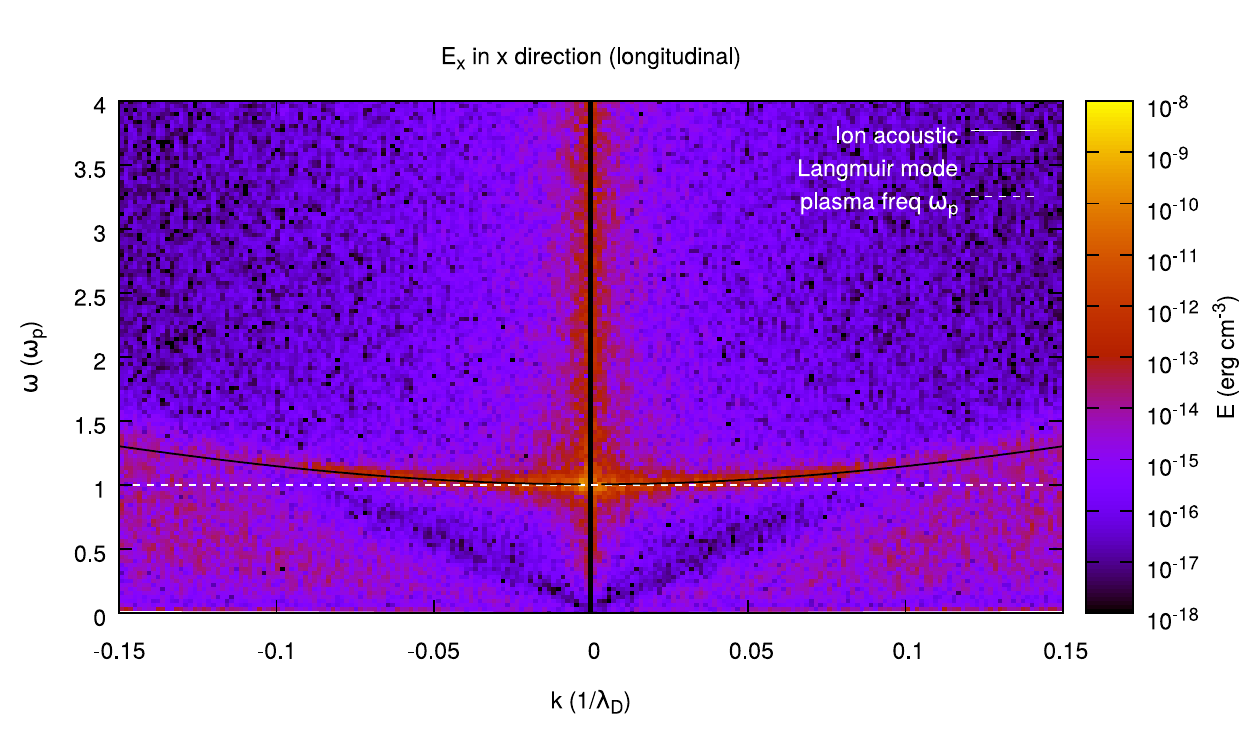}
\caption{
The distribution of energy density in the fluctuation of the electric field
in the Fourier domain compares well with the theoretical dispersion relation of
Langmuir waves. The gap in the spectrum at $k_\mathrm{x} = 0$ is an artifact of
the spectral solver. At low frequencies ion acoustic waves are also visible.
}
\label{fig:langmuir}
\end{figure*}

The expected dispersion relation of Langmuir waves is given by (see e.g.
\cite{Koskinen_2011}):
\begin{equation}
\omega = \omega_\mathrm{p} \, \sqrt{1 + 3 \, k^2 \, \lambda^2_\mathrm{D}} \quad .
\label{eqn:langmuir}
\end{equation}

Figure~\ref{fig:langmuir} shows an example that was plotted from the output of
a simulation using the \texttt{disp.json} configuration file that is supplied
along with the code as well as the expected dispersion curve.
The plot does not extend out to the maximal $k_\mathrm{x}$ that is present in
the simulation as the Langmuir mode is increasingly damped. In the range of
wavelengths where the Langmuir mode is present a sharp increase in energy
density can be seen along the dispersion curve expected from theory. At very
small $k_\mathrm{x}$ the wave frequency matches the plasma frequency from the
configuration file as expected. The gap at $k_\mathrm{x} = 0$ is an artifact of
the spectral solver, as explained in section~\ref{subsec:design}. At larger
$k_\mathrm{x}$ the wave frequency increases as expected. This indicates that
the effects of the finite temperature of the plasma or alternatively the
thermal distribution of particle speeds is correctly represented by the code.
At frequencies much below the plasma frequency -- barely resolved in this
simulation -- ion acoustic waves can be seen.

This test case is sufficiently fast to run and check that it is worthwhile to
do so after every change to the code or whenever compiling in a new
environment.

\subsection{Landau damping}
\label{subsec:landau}

The second test initializes a sinusoidal electron density fluctuation which
produces a strong Langmuir wave that then undergoes Landau damping (named after
\cite{Landau_1946} where the process was first described). The ions in this
test are assumed to act as a static and homogeneous background that compensates
the average charge density of the electrons. Since the spectral solver enforces
overall neutrality by construction, the ions don't have to be represented
through marker particles which cuts the computational effort roughly in half.
The initialization as well as some of the analysis follows the simulations in
\cite{Manfredi_1997}.

Langmuir waves are heavily damped if their wavelength is smaller than the Debye
length or close to it. Therefore a box of length $L = 5\pi \,
\lambda_\mathrm{D}$ was chosen and the longest wavelength mode is excited which
has $k = 0.4\, \lambda_\mathrm{D}^{-1}$. Sufficient resolution of phase space is
necessary to avoid dissipation through numerical effects that could mimic the
dissipation through Landau damping. For this reason $N_\mathrm{x} = 512$ grid
cells are used to represent the length of the simulation domain. This
implicitly determines $\Delta x$. In the configuration file
\texttt{rescale\_dx} is set accordingly.

Following Eq. \eqref{eqn:langmuir} the wave frequency $\omega$ is $1.22\,
\omega_p$. To resolve the oscillation with a sufficient number of steps the
time step is chosen to be $\Delta t = 0.1\, \omega_p^{-1}$. This is well below
the stability limit of the code and configured through the use of
\texttt{rescale\_dt} in the configuration file.

The cutoff scale in velocity is $v_\mathrm{range} = 10 \, v_\mathrm{th,e}$,
somewhat larger than in the Eulerian code used by \cite{Manfredi_1997}.
Consequently the number of grid cells in velocity direction is also lightly
larger at $N_\mathrm{vx} = 6600$. The actual value of $v_\mathrm{th,e}$ does
not matter as the simulation is rescaled to the corresponding Debye length, as
long as the assumption of non-relativistic motion is not violated.

The initial electron phase space distribution is given by:
\begin{equation}
f_0\left(x, v_\mathrm{x}\right) = \left(1 + \alpha \cdot \cos{\left(k\,x\right)}\right) \cdot \frac{1}{\sqrt{2 \pi}} \, \exp\left(- \frac{v^2_\mathrm{x}}{2 \, v^2_\mathrm{th,e}}\right) \quad .
\end{equation}
The parameter $\alpha$ gives the strength of the initial fluctuation and is set
to five percent in this test. This is sufficiently small to get linear Landau
damping initially, but large enough to get oscillation of the amplitude of the
electric field at later times.

Initially the wave with $k = 0.4\, \lambda_\mathrm{D}^{-1}$ and $\omega =
1.22\, \omega_p$ propagates with a phase velocity of $\omega / k = 3.04 \,
v_\mathrm{th,e}$. This is in the tail of the thermal distribution but well
below the cutoff $v_\mathrm{range}$. Particles that travel at the same speed as
the wave undergo a resonant interaction with the wave and gain energy. Details
of this process can be found in \cite{Dawson_1961,ONeil_1965} or many
textbooks. Overall the interaction with the particles leads to a decay of the
wave amplitude with a damping rate $\gamma$ that is given by
\begin{equation}
\gamma = \frac{\pi \, \omega \, \omega^2_\mathrm{p}}{2 \, k^2} \, f'\left(\frac{\omega}{k}\right) \quad ,
\end{equation}
where $f'$ is the derivative of the distribution function with respect to
velocity. Inserting a Maxwellian as distribution function we find that the
phase space density decreases with velocity and the derivative is negative as
expected. Using the parameters of the simulation we expect $\gamma = -0.072 \,
\omega_\mathrm{p}^{-1}$.

\begin{figure}[htb]
\includegraphics[width=\linewidth]{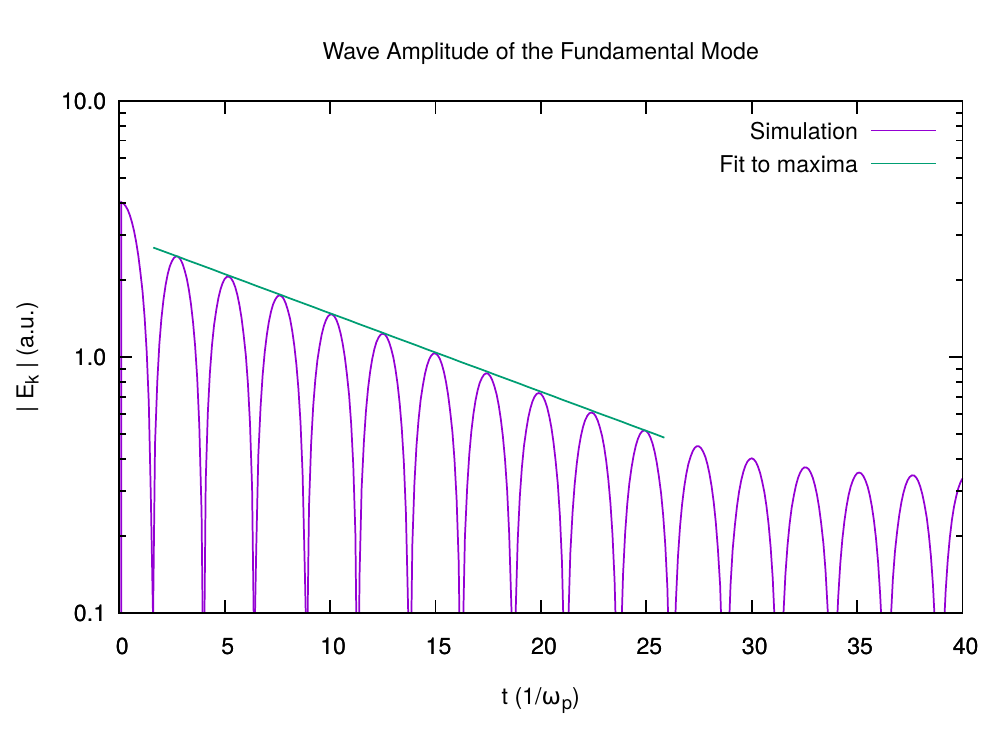}
\caption{
Short term evolution of the amplitude of the long wavelength perturbation in
the electric field as well as a fit describing linear Landau damping.
}
\label{fig:landau_test_short}
\end{figure}

The evolution at time up to $40\,\omega^{-1}_\mathrm{p}$ can be seen in
figure~\ref{fig:landau_test_short} and the expected linear Landau damping can be
recovered in the time between $1.5\,\omega^{-1}_\mathrm{p}$ and
$25\,\omega^{-1}_\mathrm{p}$. During this time the amplitude peaks of the
initially perturbed mode of the electric field can be described well by
$E_\mathrm{k} \propto \exp{\left(\gamma\,\omega_\mathrm{p}\,t\right)}$. The fit
to the simulated data finds a damping rate $\gamma$ of
$-0.070\,\omega_\mathrm{p}^{-1}$ which is close to the value that is expected from
theory. This is -- just like the correct reproduction of the dispersion curve
of the Langmuir waves itself -- a strong indication that the code simulates the
electron dynamics correctly.

Much more interesting is the study of the late time behavior of the system past
the limit of linear Landau damping. In this case the wave has modified the
distribution functions substantially close to the resonant speed. Shallow hole
like structures form in phase space and particles can get trapped there. After
one complete trapped orbit some of the original density perturbation is
restored and with it the resulting electric field.

The time dependent modifications in \cite{ONeil_1965} in principle describe
this effect and one could integrate them over time to get the amplitude of the
electric field. The much more straight forward way is to simulate the
self-consistent evolution of electron density and electric field. This
simulation requires a very low level of noise in the simulation, very few
numerical collisions that lead to dissipation and a very good representation of
fine structures in phase space. Using the parameters given above, the VHS code
can achieve this, whereas particle-in-cell codes have a hard time, given their
inherent shot noise.

\begin{figure}[htb]
\includegraphics[width=\linewidth]{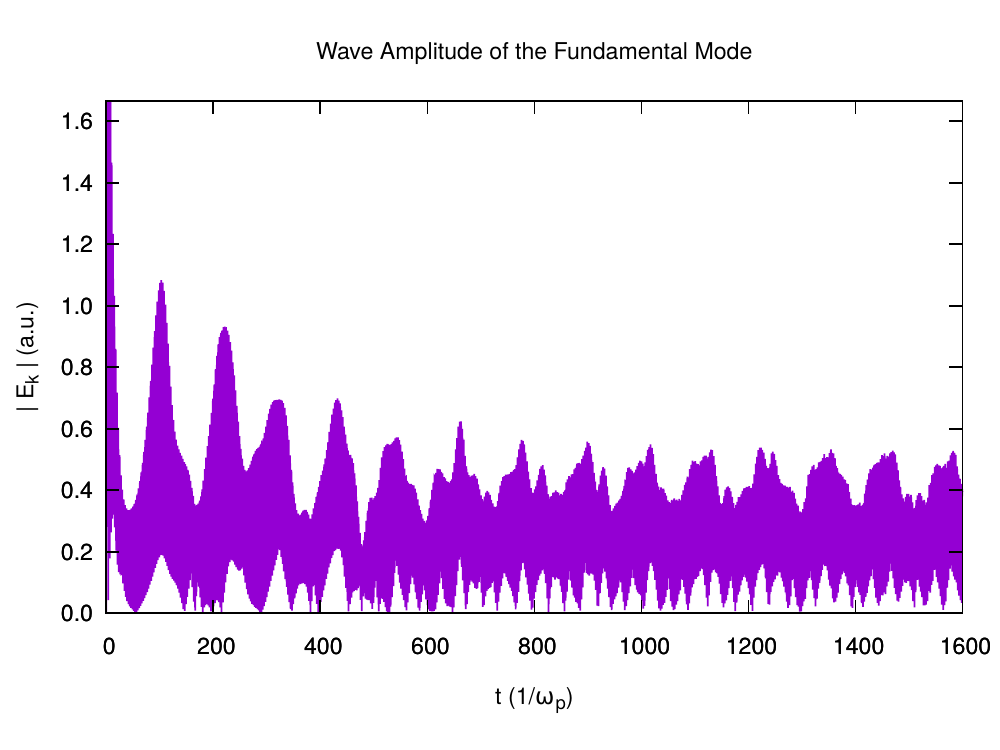}
\caption{
The long term evolution of the amplitude of the long wavelength perturbation in
the electric field shows amplitude oscillations and a non vanishing late time amplitude.
}
\label{fig:landau_test_long}
\end{figure}

Figure~\ref{fig:landau_test_long} shows the amplitude of the perturbed mode of
the electric field. To produce this plot the code stores the electric field
every five time steps. In post-processing the electric field is Fourier
transformed and the longest wavelength mode is extracted. The amplitude of this
mode can then be plotted over time.

As expected from theory and seen in the reference simulations, the electric
field does not continue to decay away but shows peaks in
amplitude caused by trapping oscillations. These peaks blur out and lose
amplitude, but retain finite values even at very late times. This shows that
fine structures in phase space are retained over long times and can form larger
coherent structures. This is only possible because the code offers good
resolution in phase space as well as a low level of numerical diffusion.

\subsection{Current-Driven Ion-Acoustic Instability}
\label{subsec:cdiai}

This third test problem uses two initially homogeneous populations (protons and
electrons) with a relative drift between each other. The setup is chosen such
that the center of mass is at rest, but a relative drift and thus a net current
in the system exist. This net current acts as a source of free energy for the
Current-Driven Ion-Acoustic instability that converts kinetic energy of the
streaming motion into electric energy of ion-acoustic waves.

For a wave mode at real frequency $\omega$ the growth rate $\gamma$ can be
found in \cite{Fitzpatrick_2014} and is given by
\begin{equation}
	\frac{\gamma}{\omega} = - \sqrt{\frac{\pi}{8}} \left[\left(\frac{m_\mathrm{e}}{m_\mathrm{i}}\right)^{1/2} \left(1-\frac{u}{c_\mathrm{s}}\right) + \left(\frac{T_\mathrm{e}}{T_\mathrm{i}}\right)^{3/2} \exp{\left(-\frac{T_\mathrm{e}}{2 T_\mathrm{i}}\right)}\right] \quad ,
	\label{eqn:iagrowth}
\end{equation}
where $c_\mathrm{s} = \sqrt{k_\mathrm{B} T_\mathrm{e}/m_\mathrm{i}}$ is the
sound speed and $u$ is the speed of the relative drift. The growth rate
obviously strongly depends on the mass ratio $m_\mathrm{i}/m_\mathrm{e}$ of
ions and electrons as well as their temperature ratio
$T_\mathrm{e}/T_\mathrm{i}$. Same as in the first test problem, the natural
value for the mass ratio is used, but unlike there the species are not in
thermal equilibrium. Equation~\eqref{eqn:iagrowth} is only valid in the regime
$T_\mathrm{e}/T_\mathrm{i} \gg 1$, where Landau damping is sufficiently weak.
A temperature ratio of 16 was chosen, as it is sufficiently large to make this
approximation valid and is convenient to set up numerically.

\begin{figure}[htb]
\includegraphics[width=\linewidth]{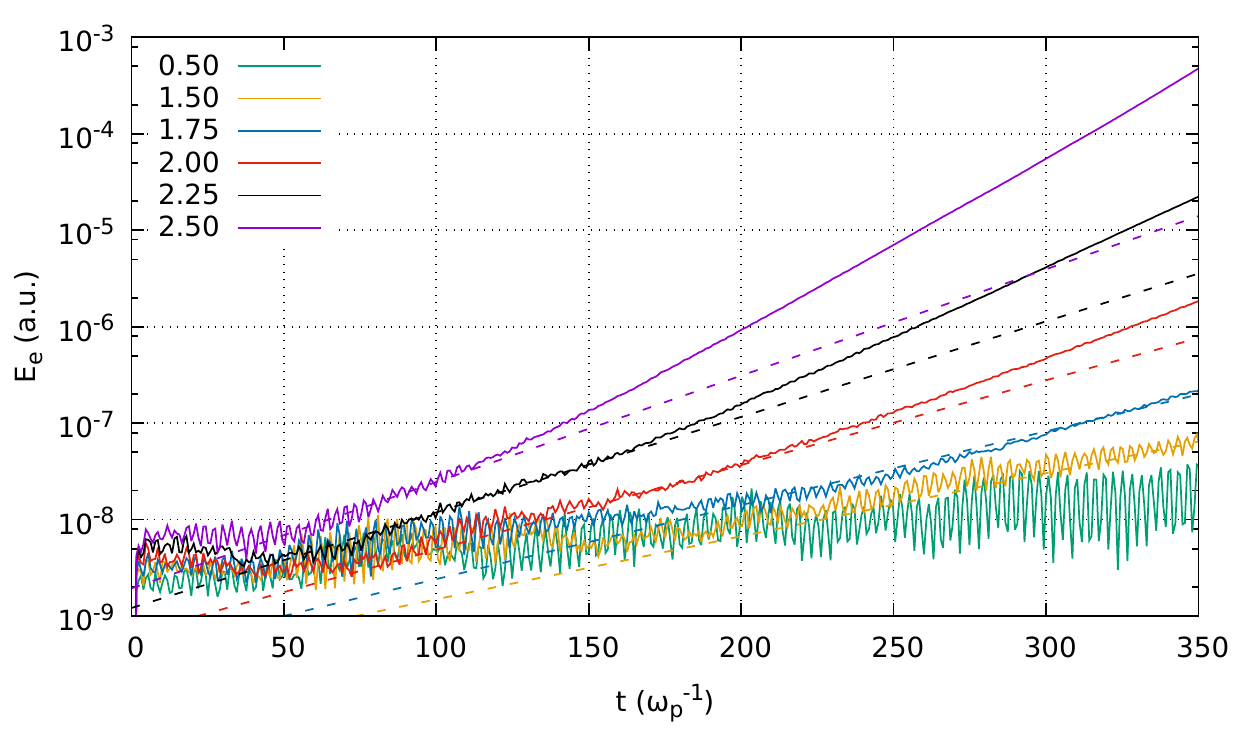}
\caption{
Growth of the energy in the electrostatic field. The curves are labeled by the
relative drift speed $u$ as multiples of the electron thermal velocity
$v_\mathrm{th,e}$. Solid lines give the result from the simulation, dashed
lines show exponential growth with the theoretically predicted growth rate,
adjusted in initial amplitude to match the simulation results during the
initial growth phase (defined here as the energy range $5\cdot10^{-9} - 5\cdot10^{-8}$
a.u.).
}
\label{fig:iagrowth}
\end{figure}

It can be seen that $\gamma$ is only positive, indicating wave growth, when the
drift speed $u$ is sufficiently large. Alternatively this is also visible from
figure~\ref{fig:iagrowth} that shows the evolution of the energy in the
electrostatic field for different relative drift speeds.

The figure shows that for a small initial drift -- below the threshold -- there
is an exchange in energy between the particles and the electric field, but
the latter doesn't show a long term growth. When the drift speed is increased,
the growth rate increases. The dashed lines also included in the plot indicate
exponential growth with the rate given by equation~\eqref{eqn:iagrowth}. The only
adjustment that was made to match the theoretical prediction to the simulation
results is setting the initial amplitude such that the theoretical curve
matches the simulation result in the first decade of energy above the noise
floor.

The figure also shows that once the energy in the ion acoustic waves has grown
by more than a factor of 20, non-linear effects set in. Interestingly this does
not reduce the growth rate as it is the case in many other instabilities, but
actually increases the growth rate. This has also been observed in other
studies of this instability, such as \cite{Lesur_2014}. The simulation there,
however, used a tiny mass ratio of four which eliminates the possibility of a
direct comparison.

\begin{figure}[htb]
\includegraphics[width=0.62\linewidth]{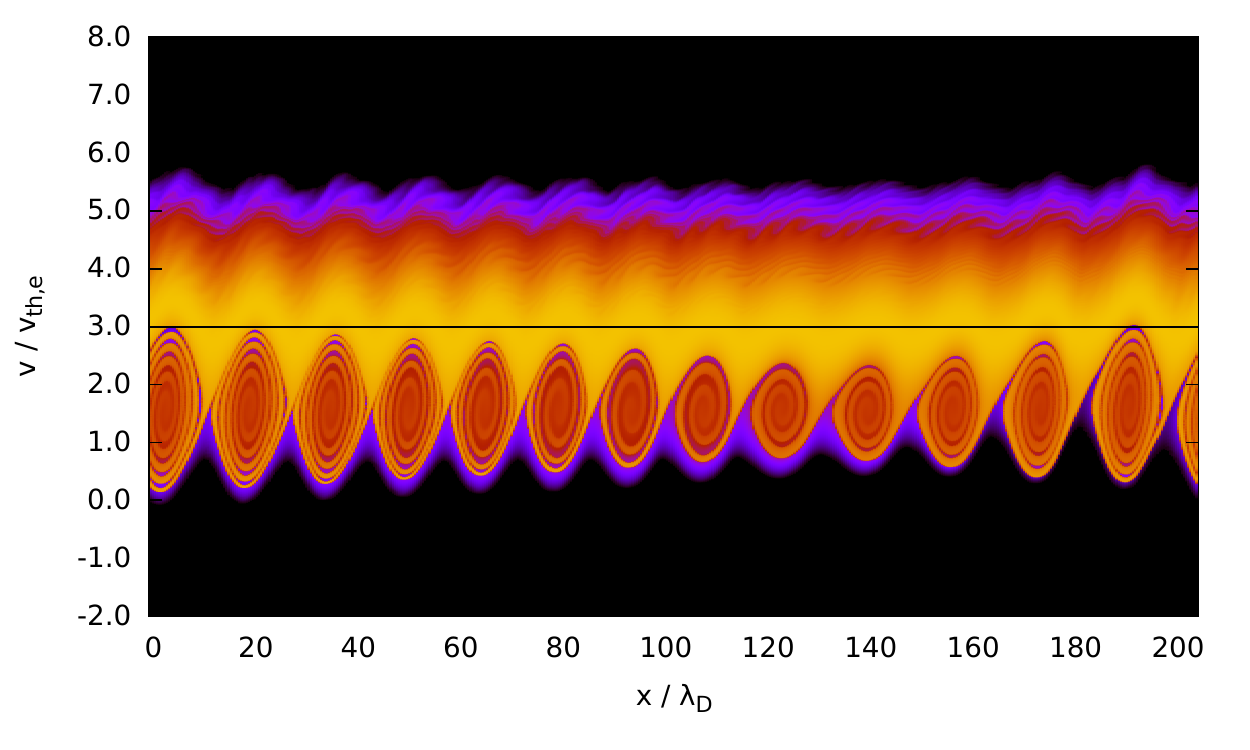}%
\includegraphics[width=0.37\linewidth]{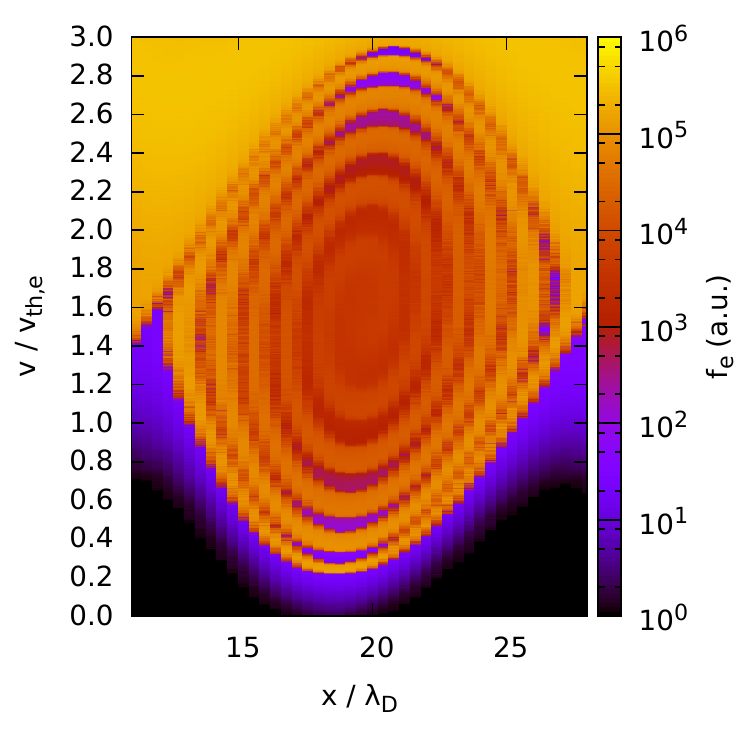}%
\caption{This plot shows the phase space density of electrons during the
non-linear stage of the current-driven ion-acoustic instability. Initially the
electrons drift at three times their thermal speed relative to the ions. At $t
= 392 \, \omega_\mathrm{p}^{-1}$ the initial homogeneous electron distribution
function develops the structures shown here.}
\label{fig:iaphasespace}
\end{figure}

Figure~\ref{fig:iaphasespace} a) shows that the phase space of electrons
develops strong deviations from the initial drifting Maxwellian. These are
especially pronounced at lower velocities (as seen from the rest frame of the
center of mass) where they interact more efficiently with the ions. Part b) of
the figure shows a zoom in of a part of phase space to illustrate the sharp
change in phase space density on small scales in both space and velocity.
Within less than one Debye length or less than one thermal speed the phase
space density changes by a factor of $10^5$. This is well represented here
using the VHS method, but is a hard problem for a particle-in-cell code.

\subsection{Electron Bernstein Modes}
\label{subsec:bernstein}

To test the 1d3v version of the code a test problem is needed that requires all
three velocity components to be resolved. For simulations that only cover one
dimension in space this often happens when a background magnetic field is
present and the gyration of particles around it have to be considered. To
demonstrate the correct gyration as well as the interaction through the
electrostatic field we consider the case of electron Bernstein modes in a
moderately warm ($v_\mathrm{th,i} = 1/10\, v_\mathrm{A}$) plasma. This test case has been used in \cite{Kilian_2017} to
test a number of different plasma models.

\begin{figure*}[hbt]
\includegraphics[width=\textwidth]{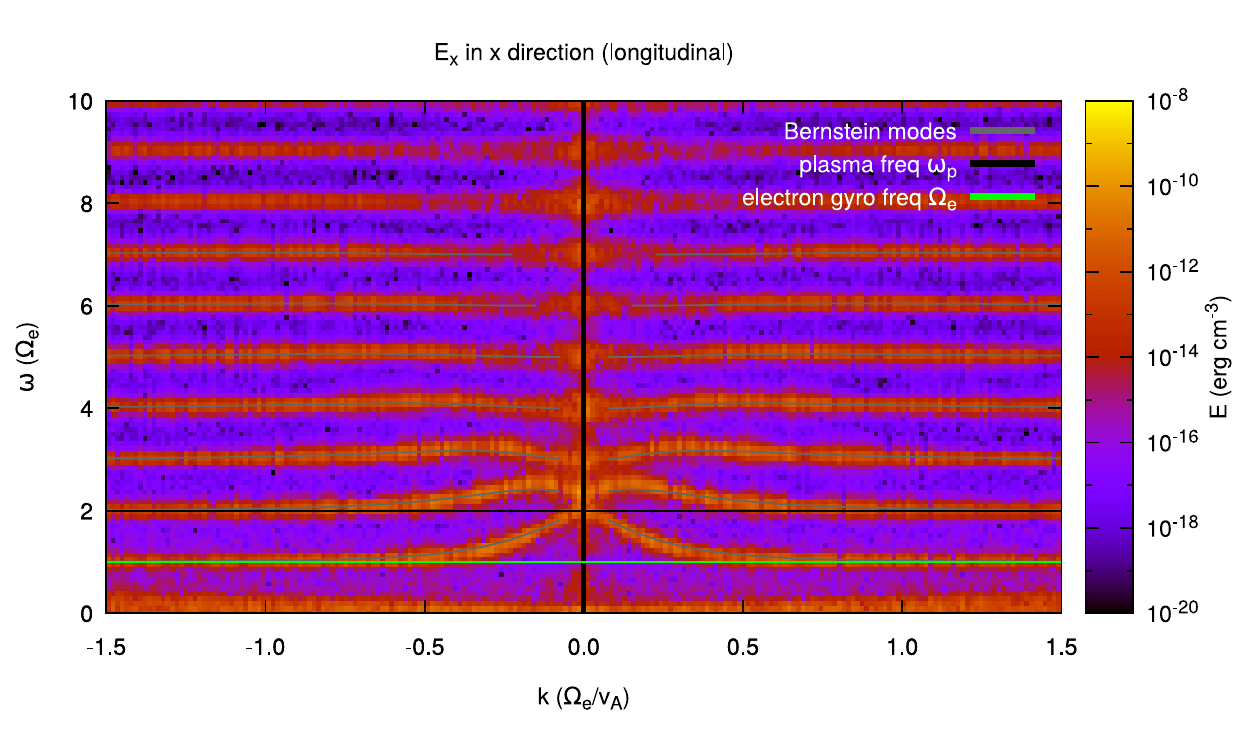}
\caption{
The distribution of energy density in the fluctuation of the electric field
in the Fourier domain compares well with the theoretical dispersion relation of
electron Bernstein waves. The gap in the spectrum at $k_\mathrm{x} = 0$ is an artifact of
the spectral solver. At low frequencies ion acoustic waves are also visible.
}
\label{fig:bernstein}
\end{figure*}

Figure~\ref{fig:bernstein} shows the result of a simulation using the
\texttt{bernstein\_3d.json} file that is supplied with the simulation code. The
distribution of the energy density in the Fourier domain matches the expected
behavior of electron Bernstein waves well. The simulation was performed with
128 cells in each of the three velocity dimensions. Consequently it requires
about 700GB of RAM. This large amount of main memory required in 1d3v
simulations was the initial reason to select parallelization using MPI as that
allows simulations using more memory than available on each individual machine
the researcher has access to. It is however also a useful performance feature
for simulations with fewer resolved dimensions.




\section{Conclusion}
\label{sec:conclusion}

This paper presents the first freely available code in the Computer Physics
Communications Program Library for Vlasov-Hybrid-Simulations. The code performs well
and is parallelized using MPI and scales well to several hundred parallel
tasks. Output is handled using parallel HDF5 and can be configured in several
levels of detail. The configuration is read at startup from JSON files that can
serve as a binding element in the pipeline from simulation design to
post-processing and plotting. Despite all these advanced features the code is
2000 lines of code in 15 files. It is well structured and aims to be
readable.

As such it offers a good introduction to the VHS method that is less well-known
than other semi-Lagrangian methods such as particle-in-cell. Despite the lower
adoption in the community the method offers several advantages. It allows to
efficiently represent small changes in phase space density. The simulations
have a very low numerical noise level and very low diffusion in phase space.
This makes VHS the method of choice to study late time effects or processes
requiring a large dynamic range above the numerical noise floor. A possible
additional advantage of the method that is not realized in the code for reasons
of simplicity is the ability to represent boundaries in phase space that are
hard to implement in particle-in-cell codes.

\appendix

\section{Acknowledgments}
The authors want to thank Mehdi Jenab for useful discussions.
%
A description of the design and implementation of FFTW3 library can be found in
\cite{Frigo_2005}.
The HDF5 library was developed by The HDF Group and by the National Center for
Supercomputing Applications at the University of Illinois at Urbana-Champaign.
%
%
Design and implementation of the MPI standard by OpenMPI is described in
\cite{Gabriel_2004}.
This work is based upon research supported by the National Research Foundation
and Department of Science and Technology. Any opinion, findings and conclusions
or recommendations expressed in this material are those of the authors and
therefore the NRF and DST do not accept any liability in regard thereto.

\section{References}
\bibliographystyle{elsarticle-num}
\bibliography{paper}

\begin{thebibliography}{10}
\expandafter\ifx\csname url\endcsname\relax
  \def\url#1{\texttt{#1}}\fi
\expandafter\ifx\csname urlprefix\endcsname\relax\def\urlprefix{URL }\fi
\expandafter\ifx\csname href\endcsname\relax
  \def\href#1#2{#2} \def\path#1{#1}\fi

\bibitem{Nunn_1993}
D.~Nunn, A novel technique for the numerical simulation of hot collision-free
  plasma; {Vlasov Hybrid Simulation}, Journal of Computational Physics 108~(1)
  (1993) 180--196.
\newblock \href {http://dx.doi.org/10.1006/jcph.1993.1173}
  {\path{doi:10.1006/jcph.1993.1173}}.

\bibitem{Dawson_1983}
J.~M. Dawson, Particle simulation of plasmas, Reviews of Modern Physics 55
  (1983) 403--447.
\newblock \href {http://dx.doi.org/10.1103/RevModPhys.55.403}
  {\path{doi:10.1103/RevModPhys.55.403}}.

\bibitem{Cheng_1976}
C.~Cheng, G.~Knorr, The integration of the {Vlasov} equation in configuration
  space, Journal of Computational Physics 22~(3) (1976) 330--351.
\newblock \href
  {http://dx.doi.org/https://doi.org/10.1016/0021-9991(76)90053-X}
  {\path{doi:https://doi.org/10.1016/0021-9991(76)90053-X}}.

\bibitem{Alfthan_2014}
S.~von Alfthan, D.~Pokhotelov, Y.~Kempf, S.~Hoilijoki, I.~Honkonen,
  A.~Sandroos, M.~Palmroth, Vlasiator: First global hybrid-{Vlasov} simulations
  of earth's foreshock and magnetosheath, Journal of Atmospheric and
  Solar-Terrestrial Physics 120 (2014) 24--35.
\newblock \href {http://dx.doi.org/https://doi.org/10.1016/j.jastp.2014.08.012}
  {\path{doi:https://doi.org/10.1016/j.jastp.2014.08.012}}.

\bibitem{Lynden_1967}
D.~Lynden-Bell, {Statistical mechanics of violent relaxation in stellar
  systems}, Monthly Notices of the Royal Astronomical Society 136 (1967) 101.
\newblock \href {http://dx.doi.org/10.1093/mnras/136.1.101}
  {\path{doi:10.1093/mnras/136.1.101}}.

\bibitem{Denavit_1972}
J.~Denavit, Numerical simulation of plasmas with periodic smoothing in phase
  space, Journal of Computational Physics 9~(1) (1972) 75--98.
\newblock \href
  {http://dx.doi.org/https://doi.org/10.1016/0021-9991(72)90037-X}
  {\path{doi:https://doi.org/10.1016/0021-9991(72)90037-X}}.

\bibitem{Symon_1970}
K.~Symon, D.~Marshall, K.~Li, Bit-pushing and distribution-pushing techniques
  for the solution of the {Vlasov} equation, in: J.~Boris, R.~Shanny (Eds.),
  Proceedings of the Fourth Conference on the Numerical Simulation of Plasmas,
  Washington DC, Naval Research Laboratory, Washington DC, 1970, pp. 68--125.

\bibitem{Joyce_1971}
G.~Joyce, G.~Knorr, H.~K. Meier, Numerical integration methods of the {Vlasov}
  equation, Journal of Computational Physics 8~(1) (1971) 53--63.
\newblock \href
  {http://dx.doi.org/https://doi.org/10.1016/0021-9991(71)90034-9}
  {\path{doi:https://doi.org/10.1016/0021-9991(71)90034-9}}.

\bibitem{Shoucri_1978}
M.~M. Shoucri, R.~R. Gagn{\'e}, Splitting schemes for the numerical solution of
  a two-dimensional {Vlasov} equation, Journal of Computational Physics 27~(3)
  (1978) 315--322.
\newblock \href
  {http://dx.doi.org/https://doi.org/10.1016/0021-9991(78)90013-X}
  {\path{doi:https://doi.org/10.1016/0021-9991(78)90013-X}}.

\bibitem{Nakamura_Yabe_1999}
T.~Nakamura, T.~Yabe, Cubic interpolated propagation scheme for solving the
  hyper-dimensional {Vlasov}-{Poisson} equation in phase space, Computer
  Physics Communications 120~(2) (1999) 122--154.
\newblock \href
  {http://dx.doi.org/https://doi.org/10.1016/S0010-4655(99)00247-7}
  {\path{doi:https://doi.org/10.1016/S0010-4655(99)00247-7}}.

\bibitem{Knorr_1963}
G.~Knorr, {Zur L{\"o}sung der nicht-linearen Vlasov-Gleichung}, Zeitschrift
  Naturforschung Teil A 18 (1963) 1304--1315.
\newblock \href {http://dx.doi.org/10.1515/zna-1963-1209}
  {\path{doi:10.1515/zna-1963-1209}}.

\bibitem{Denavit_1971}
J.~Denavit, W.~L. Kruer, Comparison of numerical solutions of the {Vlasov}
  equation with particle simulations of collisionless plasmas, The Physics of
  Fluids 14~(8) (1971) 1782--1791.
\newblock \href {http://dx.doi.org/10.1063/1.1693676}
  {\path{doi:10.1063/1.1693676}}.

\bibitem{Klimas_1987}
A.~J. {Klimas}, {A method for overcoming the velocity space filamentation
  problem in collisionless plasma model solutions}, Journal of Computational
  Physics 68 (1987) 202--226.
\newblock \href {http://dx.doi.org/10.1016/0021-9991(87)90052-0}
  {\path{doi:10.1016/0021-9991(87)90052-0}}.

\bibitem{Klimas_1994}
A.~Klimas, W.~Farrell, A splitting algorithm for {Vlasov} simulation with
  filamentation filtration, Journal of Computational Physics 110~(1) (1994)
  150--163.
\newblock \href {http://dx.doi.org/https://doi.org/10.1006/jcph.1994.1011}
  {\path{doi:https://doi.org/10.1006/jcph.1994.1011}}.

\bibitem{Eliasson_2002}
B.~Eliasson, Outflow boundary conditions for the fourier transformed
  two-dimensional vlasov equation, Journal of Computational Physics 181~(1)
  (2002) 98--125.
\newblock \href {http://dx.doi.org/https://doi.org/10.1006/jcph.2002.7121}
  {\path{doi:https://doi.org/10.1006/jcph.2002.7121}}.

\bibitem{Boris_1976}
J.~Boris, D.~Book, Flux-corrected transport. {III}. minimal-error {FCT}
  algorithms, Journal of Computational Physics 20~(4) (1976) 397--431.
\newblock \href
  {http://dx.doi.org/https://doi.org/10.1016/0021-9991(76)90091-7}
  {\path{doi:https://doi.org/10.1016/0021-9991(76)90091-7}}.

\bibitem{Fijalkow_1999}
E.~Fijalkow, A numerical solution to the {Vlasov} equation, Computer Physics
  Communications 116~(2) (1999) 319--328.
\newblock \href
  {http://dx.doi.org/https://doi.org/10.1016/S0010-4655(98)00146-5}
  {\path{doi:https://doi.org/10.1016/S0010-4655(98)00146-5}}.

\bibitem{Filbet_2003}
F.~Filbet, E.~Sonnendr{\"u}cker, Comparison of {Eulerian} {Vlasov} solvers,
  Computer Physics Communications 150~(3) (2003) 247--266.
\newblock \href
  {http://dx.doi.org/https://doi.org/10.1016/S0010-4655(02)00694-X}
  {\path{doi:https://doi.org/10.1016/S0010-4655(02)00694-X}}.

\bibitem{Jenko_2000}
F.~Jenko, W.~Dorland, M.~Kotschenreuther, B.~N. Rogers, Electron temperature
  gradient driven turbulence, Physics of Plasmas 7~(5) (2000) 1904--1910.
\newblock \href {http://dx.doi.org/10.1063/1.874014}
  {\path{doi:10.1063/1.874014}}.

\bibitem{Sonnendrücker_1999}
E.~Sonnendr{\"u}cker, J.~Roche, P.~Bertrand, A.~Ghizzo, The semi-lagrangian
  method for the numerical resolution of the vlasov equation, Journal of
  Computational Physics 149~(2) (1999) 201--220.
\newblock \href {http://dx.doi.org/https://doi.org/10.1006/jcph.1998.6148}
  {\path{doi:https://doi.org/10.1006/jcph.1998.6148}}.

\bibitem{Barnes_1986}
J.~Barnes, P.~Hut, {A hierarchical O(N log N) force-calculation algorithm},
  Nature 324 (1986) 446--449.
\newblock \href {http://dx.doi.org/10.1038/324446a0}
  {\path{doi:10.1038/324446a0}}.

\bibitem{Dubinski_1996}
J.~Dubinski, A parallel tree code, New Astronomy 1~(2) (1996) 133--147.
\newblock \href {http://dx.doi.org/10.1016/S1384-1076(96)00009-7}
  {\path{doi:10.1016/S1384-1076(96)00009-7}}.

\bibitem{Gibbon_2004}
P.~Gibbon, F.~N. Beg, E.~L. Clark, R.~G. Evans, M.~Zepf, Tree-code simulations
  of proton acceleration from laser-irradiated wire targets, Physics of Plasmas
  11~(8) (2004) 4032--4040.
\newblock \href {http://dx.doi.org/10.1063/1.1767096}
  {\path{doi:10.1063/1.1767096}}.

\bibitem{Winkel_2012}
M.~Winkel, R.~Speck, H.~H{\"u}bner, L.~Arnold, R.~Krause, P.~Gibbon, A
  massively parallel, multi-disciplinary barnes–hut tree code for
  extreme-scale n-body simulations, Computer Physics Communications 183~(4)
  (2012) 880--889.
\newblock \href {http://dx.doi.org/10.1016/j.cpc.2011.12.013}
  {\path{doi:10.1016/j.cpc.2011.12.013}}.

\bibitem{Birdsall_2005}
C.~K. {Birdsall}, A.~B. {Langdon}, Plasma physics via computer simulation, 1st
  Edition, New York: Taylor and Francis, 2005.

\bibitem{Dawson_1962}
J.~Dawson, One-dimensional plasma model, The Physics of Fluids 5~(4) (1962)
  445--459.
\newblock \href {http://dx.doi.org/10.1063/1.1706638}
  {\path{doi:10.1063/1.1706638}}.

\bibitem{Hockney_1966}
R.~W. Hockney, Computer experiment of anomalous diffusion, The Physics of
  Fluids 9~(9) (1966) 1826--1835.
\newblock \href {http://dx.doi.org/10.1063/1.1761939}
  {\path{doi:10.1063/1.1761939}}.

\bibitem{Birdsall_1969}
C.~K. Birdsall, D.~Fuss, Clouds-in-clouds, clouds-in-cells physics for
  many-body plasma simulation, Journal of Computational Physics 3~(4) (1969)
  494--511.
\newblock \href
  {http://dx.doi.org/https://doi.org/10.1016/0021-9991(69)90058-8}
  {\path{doi:https://doi.org/10.1016/0021-9991(69)90058-8}}.

\bibitem{Friedman_1992}
A.~Friedman, D.~P. Grote, I.~Haber, Three-dimensional particle simulation of
  heavy-ion fusion beams, Physics of Fluids B: Plasma Physics 4~(7) (1992)
  2203--2210.
\newblock \href {http://dx.doi.org/10.1063/1.860024}
  {\path{doi:10.1063/1.860024}}.

\bibitem{Fonseca_2002}
R.~A. Fonseca, L.~O. Silva, F.~S. Tsung, V.~K. Decyk, W.~Lu, C.~Ren, W.~B.
  Mori, S.~Deng, S.~Lee, T.~Katsouleas, J.~C. Adam, {OSIRIS}: A
  three-dimensional, fully relativistic particle in cell code for modeling
  plasma based accelerators, in: P.~M.~A. Sloot, A.~G. Hoekstra, C.~J.~K. Tan,
  J.~J. Dongarra (Eds.), Computational Science - {ICCS} 2002, Springer Berlin
  Heidelberg, Berlin, Heidelberg, 2002, pp. 342--351.

\bibitem{Arber_2015}
T.~D. Arber, K.~Bennett, C.~S. Brady, A.~Lawrence-Douglas, M.~G. Ramsay, N.~J.
  Sircombe, P.~Gillies, R.~G. Evans, H.~Schmitz, A.~R. Bell, C.~P. Ridgers,
  \href{http://stacks.iop.org/0741-3335/57/i=11/a=113001}{Contemporary
  particle-in-cell approach to laser-plasma modelling}, Plasma Physics and
  Controlled Fusion 57~(11) (2015) 113001.
\newline\urlprefix\url{http://stacks.iop.org/0741-3335/57/i=11/a=113001}

\bibitem{Boris_1970}
J.~P. Boris, Relativistic plasma simulation---optimization of a hybrid code,
  in: J.~Boris, R.~Shanny (Eds.), Proceedings of the Fourth Conference on the
  Numerical Simulation of Plasmas, Washington DC, Naval Research Laboratory,
  Washington DC, 1970, pp. 3--67.

\bibitem{Nunn_1990}
D.~Nunn, The numerical simulation of vlf nonlinear wave-particle interactions
  in collision-free plasmas using the {Vlasov} hybrid simulation technique,
  Computer Physics Communications 60~(1) (1990) 1--25.
\newblock \href
  {http://dx.doi.org/https://doi.org/10.1016/0010-4655(90)90074-B}
  {\path{doi:https://doi.org/10.1016/0010-4655(90)90074-B}}.

\bibitem{Kazeminezhad_2003}
F.~Kazeminezhad, S.~Kuhn, A.~Tavakoli, {Vlasov} model using kinetic phase point
  trajectories, Phys. Rev. E 67 (2003) 026704.
\newblock \href {http://dx.doi.org/10.1103/PhysRevE.67.026704}
  {\path{doi:10.1103/PhysRevE.67.026704}}.

\bibitem{Powell_1999}
K.~G. Powell, P.~L. Roe, T.~J. Linde, T.~I. Gombosi, D.~L.~D. Zeeuw, A
  solution-adaptive upwind scheme for ideal magnetohydrodynamics, Journal of
  Computational Physics 154~(2) (1999) 284--309.
\newblock \href {http://dx.doi.org/https://doi.org/10.1006/jcph.1999.6299}
  {\path{doi:https://doi.org/10.1006/jcph.1999.6299}}.

\bibitem{Fryxell_2000}
B.~Fryxell, K.~Olson, P.~Ricker, F.~X. Timmes, M.~Zingale, D.~Q. Lamb,
  P.~MacNeice, R.~Rosner, J.~W. Truran, H.~Tufo, Flash: An adaptive mesh
  hydrodynamics code for modeling astrophysical thermonuclear flashes, The
  Astrophysical Journal Supplement Series 131~(1) (2000) 273.

\bibitem{Brandenburg_2002}
A.~Brandenburg, W.~Dobler, {Hydromagnetic turbulence in computer simulations},
  Computer Physics Communications 147 (2002) 471--475.
\newblock \href {http://dx.doi.org/10.1016/S0010-4655(02)00334-X}
  {\path{doi:10.1016/S0010-4655(02)00334-X}}.

\bibitem{Mignone_2007}
A.~Mignone, G.~Bodo, S.~Massaglia, T.~Matsakos, O.~Tesileanu, C.~Zanni,
  A.~Ferrari, Pluto: A numerical code for computational astrophysics, The
  Astrophysical Journal Supplement Series 170~(1) (2007) 228.

\bibitem{Hanasz_2010}
{Hanasz, M.}, {Kowalik, K.}, {W{\'o}lta{\'n}ski, D.}, {Paw{\l}aszek, R.},
  \href{https://doi.org/10.1051/eas/1042029}{The {PIERNIK} {MHD} code – a
  multi-fluid, non-ideal extension of the relaxing-tvd scheme ({I})}, EAS
  Publications Series 42 (2010) 275--280.
\newblock \href {http://dx.doi.org/10.1051/eas/1042029}
  {\path{doi:10.1051/eas/1042029}}.
\newline\urlprefix\url{https://doi.org/10.1051/eas/1042029}

\bibitem{Glocer_2009}
A.~Glocer, G.~T{\'o}th, Y.~Ma, T.~Gombosi, J.-C. Zhang, L.~M. Kistler,
  \href{http://dx.doi.org/10.1029/2009JA014418}{Multifluid block-adaptive-tree
  solar wind {Roe}-type upwind scheme: Magnetospheric composition and dynamics
  during geomagnetic storms - {I}nitial results}, Journal of Geophysical
  Research: Space Physics 114~(A12), a12203.
\newblock \href {http://dx.doi.org/10.1029/2009JA014418}
  {\path{doi:10.1029/2009JA014418}}.
\newline\urlprefix\url{http://dx.doi.org/10.1029/2009JA014418}

\bibitem{Leake_2012}
J.~E. Leake, V.~S. Lukin, M.~G. Linton, E.~T. Meier,
  \href{http://stacks.iop.org/0004-637X/760/i=2/a=109}{Multi-fluid simulations
  of chromospheric magnetic reconnection in a weakly ionized reacting plasma},
  The Astrophysical Journal 760~(2) (2012) 109.
\newline\urlprefix\url{http://stacks.iop.org/0004-637X/760/i=2/a=109}

\bibitem{Müller_2011}
J.~M{\"u}ller, S.~Simon, U.~Motschmann, J.~Sch{\"u}le, K.-H. Glassmeier, G.~J.
  Pringle, {A.I.K.E.F.}: Adaptive hybrid model for space plasma simulations,
  Computer Physics Communications 182~(4) (2011) 946--966.
\newblock \href {http://dx.doi.org/https://doi.org/10.1016/j.cpc.2010.12.033}
  {\path{doi:https://doi.org/10.1016/j.cpc.2010.12.033}}.

\bibitem{Kallio_2002}
E.~Kallio, P.~Janhunen, \href{http://dx.doi.org/10.1029/2001JA000090}{Ion
  escape from mars in a quasi-neutral hybrid model}, Journal of Geophysical
  Research: Space Physics 107~(A3) (2002) SIA 1--1--SIA 1--21.
\newblock \href {http://dx.doi.org/10.1029/2001JA000090}
  {\path{doi:10.1029/2001JA000090}}.
\newline\urlprefix\url{http://dx.doi.org/10.1029/2001JA000090}

\bibitem{Canosa_1972}
J.~Canosa, J.~Gazdag, J.~E. Fromm, B.~H. Armstrong, Electrostatic oscillations
  in plasmas with cutoff distributions, The Physics of Fluids 15~(12) (1972)
  2299--2305.
\newblock \href {http://dx.doi.org/10.1063/1.1693873}
  {\path{doi:10.1063/1.1693873}}.

\bibitem{Canosa_1974}
J.~Canosa, J.~Gazdag, J.~Fromm, The recurrence of the initial state in the
  numerical solution of the {Vlasov} equation, Journal of Computational Physics
  15~(1) (1974) 34--45.
\newblock \href
  {http://dx.doi.org/https://doi.org/10.1016/0021-9991(74)90067-9}
  {\path{doi:https://doi.org/10.1016/0021-9991(74)90067-9}}.

\bibitem{Manfredi_1997}
G.~Manfredi, Long-time behavior of nonlinear {Landau} damping, Physical Review
  Letters 79 (1997) 2815--2818.
\newblock \href {http://dx.doi.org/10.1103/PhysRevLett.79.2815}
  {\path{doi:10.1103/PhysRevLett.79.2815}}.

\bibitem{Pezzi_2016}
O.~Pezzi, E.~Camporeale, F.~Valentini, Collisional effects on the numerical
  recurrence in {Vlasov}-{Poisson} simulations, Physics of Plasmas 23~(2)
  (2016) 022103.
\newblock \href {http://dx.doi.org/10.1063/1.4940963}
  {\path{doi:10.1063/1.4940963}}.

\bibitem{Kilian_2012}
P.~Kilian, T.~Burkart, F.~Spanier, The influence of the mass ratio on particle
  acceleration by the filamentation instability, in: W.~E. Nagel, D.~B.
  Kr\"oner, M.~M. Resch (Eds.), High Performance Computing in Science and
  Engineering '11, Springer, Berlin Heidelberg, 2012, pp. 5--13.
\newblock \href {http://dx.doi.org/10.1007/978-3-642-23869-7}
  {\path{doi:10.1007/978-3-642-23869-7}}.

\bibitem{Sitenko_1967}
A.~G. Sitenko, Electromagnetic Fluctuations in Plasma, Academic Press, New
  York, 1967.

\bibitem{Koskinen_2011}
H.~E.~J. Koskinen, Physics of space storms from the solar surface the Earth,
  Springer ; Published in association with Praxis Pub., Berlin; London;
  Chichester, UK, 2011.
\newblock \href {http://dx.doi.org/10.1007/978-3-642-00319-6}
  {\path{doi:10.1007/978-3-642-00319-6}}.

\bibitem{Landau_1946}
L.~Landau, On the vibrations of the electronic plasma, Zhurnal Eksperimentalnoi
  i Teoreticheskoi Fiziki 16~(7) (1946) 574--586.

\bibitem{Dawson_1961}
J.~M. Dawson, On {Landau} damping, Physics of Fluids 4~(7) (1961) 869--874.
\newblock \href {http://dx.doi.org/10.1063/1.1706419}
  {\path{doi:10.1063/1.1706419}}.

\bibitem{ONeil_1965}
T.~O{\'{}}Neil, Collisionless damping of nonlinear plasma oscillations, Physics
  of Fluids 8~(12) (1965) 2255--2262.
\newblock \href {http://dx.doi.org/10.1063/1.1761193}
  {\path{doi:10.1063/1.1761193}}.

\bibitem{Fitzpatrick_2014}
R.~Fitzpatrick, Plasma Physics: An Introduction, CRC Press, 2014.

\bibitem{Lesur_2014}
M.~Lesur, P.~H. Diamond, Y.~Kosuga, Nonlinear current-driven ion-acoustic
  instability driven by phase-space structures, Plasma Physics and Controlled
  Fusion 56~(7) (2014) 075005.
\newblock \href {http://dx.doi.org/10.1088/0741-3335/56/7/075005}
  {\path{doi:10.1088/0741-3335/56/7/075005}}.

\bibitem{Kilian_2017}
P.~Kilian, P.~A. Mu{\~n}oz, C.~Schreiner, F.~Spanier, Plasma waves as a
  benchmark problem, Journal of Plasma Physics 83~(1).
\newblock \href {http://dx.doi.org/10.1017/S0022377817000149}
  {\path{doi:10.1017/S0022377817000149}}.

\bibitem{Frigo_2005}
M.~Frigo, S.~G. Johnson, The design and implementation of {FFTW3}, Proceedings
  of the IEEE 93~(2) (2005) 216--231, special issue on ``Program Generation,
  Optimization, and Platform Adaptation''.

\bibitem{Gabriel_2004}
E.~Gabriel, G.~E. Fagg, G.~Bosilca, T.~Angskun, J.~J. Dongarra, J.~M. Squyres,
  V.~Sahay, P.~Kambadur, B.~Barrett, A.~Lumsdaine, R.~H. Castain, D.~J. Daniel,
  R.~L. Graham, T.~S. Woodall, Open {MPI}: Goals, concept, and design of a next
  generation {MPI} implementation, in: Proceedings, 11th European {PVM/MPI}
  Users' Group Meeting, Budapest, Hungary, 2004, pp. 97--104.

\end{thebibliography}

\end{document}